\documentclass[a4paper,11pt]{article}
\usepackage{jheppub} 
\usepackage{lineno}
\usepackage[utf8]{inputenc}

\usepackage{amsmath}
\usepackage[sort&compress,numbers]{natbib}
\usepackage{amssymb}
\usepackage{tikz}
\RequirePackage{doi}
\usepackage{hyperref}
\usepackage{cleveref}
\usepackage{braket}
\usepackage{caption}
\usepackage{comment}
\usepackage{graphicx,url}
\usepackage{pgfplots}
\pgfplotsset{compat = 1.3}
\usepgfplotslibrary{fillbetween}

\usepackage{subcaption}
\usetikzlibrary{decorations.pathreplacing}
\hypersetup{
    colorlinks=true,       
    linkcolor=blue,          
    citecolor=blue,        
}

\newcommand{\Lcut}{L_\text{cutoff}}

\preprint{MIT-CTP/5586}

\title{\boldmath On the Causality Paradox and the Karch-Randall Braneworld as an EFT}

 \author[a,b]{Dominik Neuenfeld}
 \author[c]{and Manu Srivastava}
 \affiliation[a]{Institute for Theoretical Physics and Astrophysics, Julius-Maximilians-Universität Würzburg, Am Hubland, 97074 Würzburg, Germany}
 \affiliation[b]{Institute for Theoretical Physics, University of Amsterdam, Science Park 904, 1090 GL Amsterdam,
The Netherlands}
 \affiliation[c]{Center for Theoretical Physics, Massachusetts Institute of Technology, Cambridge, MA 02139, USA}

\emailAdd{dominik.neuenfeld@uni-wuerzburg.de}
\emailAdd{manu\textunderscore sr@mit.edu}

\abstract{Holography on cutoff surfaces can appear to be in tension with causality. For example, as argued by Omiya and Wei \cite{Omiya:2021olc}, double holography seemingly allows for superluminal signalling. In this paper we argue that the brane description of double holography should be treated as an effective theory and demonstrate that causality violations due to faster-than-light communication are not visible above the associated cutoff length scale. This suggests that end-of-the-world brane models are consistent with causality and that the apparent superluminal signalling is a UV effect. Moreover, we argue that short distance non-localities generically give rise to apparent faster-than-light propagation of signals in Anti-de Sitter space.
    
    Nonetheless, superluminal signalling indicates that the causal structure on holographic cutoff surfaces needs to be modified. We propose and study three different candidate regions that might replace the domain of dependence in the brane EFT of the Karch-Randall model. These regions are defined by unitarity on the brane, through bulk entanglement wedges and through the nice slice criterion, respectively. In all dimensions, these candidate regions exclude those parts of the domain of dependence which are affected by superluminal signalling. While all three definitions agree in two dimensions, they are different in higher dimensions.}

\begin{document}
\maketitle
\flushbottom

\section{Introduction}
The AdS/CFT correspondence \cite{Maldacena:1997re} is a powerful tool which has lead to crucial insights about quantum gravity in asymptotically AdS spacetimes. Cosmological observations suggest that we do not live in asymptotically AdS spacetime and hence there has been a lot of interest in generalizing the correspondence to more general situations in the hope of describing our universe using such a duality. One small step in this direction is to reduce the reliance of the correspondence on the nature of the spacetime boundary, e.g., by understanding holography on lower-dimensional surfaces in AdS which are at a finite distance from the bulk, e.g., \cite{Nomura:2016ikr,McGough:2016lol,Taylor:2018xcy,Grado-White:2020wlb}. 

In this paper we will focus on one important example of models that can be understood as holography at a cutoff, Karch-Randall (KR) braneworlds \cite{randall1999alternative, Karch:2000ct}. These models have recently been employed as powerful tools to understand important aspects of black hole evaporation \cite{Hawking:1975vcx, Page:1993wv, Almheiri:2019psf, Penington:2019npb}. Most notably, using holography, it has become possible to compute Page curves for certain black holes, even in higher dimensions \cite{Almheiri:2019hni, Almheiri:2019psy, Geng:2020qvw, Chen:2020uac, Chen:2020hmv} using a chain of arguments that has become known as \emph{double holography}. Double holography is essentially a bottom-up model of a holographic duality for a boundary conformal field theory (BCFT). It relates a low-dimensional BCFT with additional degrees of freedom at its boundary (\emph{the boundary description}) to a dual, higher dimensional AdS bulk which is cut off by an end-of-the-world brane (\emph{the bulk description}) \cite{Takayanagi:2011zk, Fujita:2011fp}. In addition there is a third picture (\emph{the brane description}).\footnote{The \emph{brane description} is also sometimes called the \emph{intermediate description}.} This third description arises from either only dualizing the degrees of freedom at the BCFT's boundary, or integrating out the bulk between the asymptotic boundary and the brane. It consists of a non-gravitating region (\emph{the bath}) which can freely exchange excitations through an interface with a gravitating theory on the brane world-volume. The three descriptions are illustrated in \cref{fig:three_perspectives}.

While computations of the Page curve are well established, relatively little is known about the local physics in the gravitating region of the brane perspective. In fact certain features of the brane description seem problematic or at least unusual.

One particular problem pointed out by Omiya and Wei \cite{Omiya:2021olc} is that double holography fails to satisfy assumptions of the Gao-Wald theorem \cite{Gao:2000ga} which guarantees that the time a signal takes to go between two points on the boundary is lower-bounded by the time it takes for the signal to travel along the boundary. And in fact, it is easy to show that doubly-holographic models admit faster-than-light signalling between the bath and gravitating region. This seems to suggest that faster-than-light communication is possible in the brane description. The authors dubbed this observation the \emph{causality paradox}.\footnote{This observation adds to other properties of the brane perspective, e.g., massive gravitons, which distinguish it from semi-classical Einstein gravity. And in fact it has been suggested that lessons learned in double holography, such as the island formula in higher dimensions, are properties of the specific model and should not be carried over to the familiar case of semi-classical Einstein gravity \cite{Geng:2021hlu}.}

A partial resolution to the causality paradox in the Karch-Randall setting was given in \cite{Karch:2022rvr}. There, the authors explicitly constructed a top-down example of a doubly-holographic model within supergravity and pointed out that one can at best approximately identify the geometry in the bulk picture (which is the origin of superluminal signalling) and the brane picture (where superluminal signalling becomes a problem). This suggests that the statement that communication between two points via the brane takes longer than through the bulk becomes ill-posed.

Still, this response is somewhat unsatisfactory. First, it seems that one might have to investigate on a case-by-case basis under which conditions which aspects of bottom-up models correctly capture the physics of more complicated top-down constructions, without having a way of diagnosing the reliability of computations purely in the brane description. Second, their argument does not solve the problem completely. Consider the case of a doubly-holographic model of a black hole in contact with a thermal bath. After the Page time, an island appears outside the black hole horizon \cite{Almheiri:2019yqk} which implies that the black hole interior cannot be reconstructed anymore from the defect degrees of freedom. Thus, it is unclear how the above argument \cite{Karch:2022rvr} would work in such a case.  

Here, we will approach issues with causality purely from a bottom-up perspective. In the first part of this paper we will argue that the causality paradox can be resolved even in bottom-up models in a simple and maximally conservative way: we only need to remember that the brane description is merely a low-energy effective description. And in fact, although it has been argued that the non-locality in the brane perspective should have an IR origin \cite{Omiya:2021olc}, we will demonstrate in this paper that superluminal effects do not appear on length scales larger than the length scale set by the cutoff of the effective theory.

However, despite the fact that distances below the cutoff cannot be probed in effective field theory (EFT), we can still treat the brane description as living on fixed background at leading order in $1/N$. Thus, superluminal signalling, although it cannot be resolved by any observer, is still in conflict with the causal structure of the background.\footnote{To see this, note that we could regularize the theory of length scales much smaller than the cutoff scale to associate density matrices to subregions. Due to faster-than-light signalling subregions which are causally independent based on the ``old'' speed of light might in fact not be.} We address this problem in two ways. First, we show that for an EFT on an AdS background UV non-localities generically give rise to a signal propagation speed that is faster than the naive speed of light. This again demonstrates the UV origin of the new, faster speed of light and supports the observation that the causal structure of the brane theory needs to be modified. Second, we discuss several definitions of regions that could act as possible replacements for the notion of domain of dependence and study their properties.

In particular we define the notion of \emph{unitary domain of dependence} as the largest spacetime subregion in which unitary time evolution can be used to evolve a state, the notion of \emph{holographic domain of dependence} which uses bulk entanglement wedges to define regions where a state can be obtained through the bulk, and the notion of \emph{EFT domain of the dependence} based on the nice slice criterion \cite{Polchinski:1995ta, Lowe_Thorlacius}.

Assuming that HRT surfaces \cite{Ryu:2006bv, Hubeny:2007xt} anchored on the brane compute entanglement entropies and define entanglement wedges, we can compute (bounds on) these subregions. We will show that these regions are smaller than the naive domain of dependence $\mathcal D$ and exclude all spacetime regions from $\mathcal D$ which are acausally influenced by faster-than-light signalling. This implies that all our candidate regions are compatible with faster-than-light signalling, unlike the standard domain of dependence.

In the examples we study we always observe that the EFT domain of dependence is contained in the other two candidate subregions. In other words, as long as we stay within the nice-slice limits of predictability of the effective brane theory we do not see any non-localities, breakdown of unitarity or causality violations affecting the brane physics. In summary, our results suggest that there is no contradiction to the existence of a low energy EFT on the brane as required in the brane/intermediate perspective of double holography.

The paper is organized as follows. In \cref{sec:review}, we review the Karch-Randall braneworld models, remind the reader that it comes with a cutoff scale and show how the Karch-Randall models exhibit superluminal signalling. \Cref{sec:scale_of_violation} establishes that the faster-than-light signalling cannot be resolved within the regime of validity of the effective brane theory. In \cref{sec:speed_of_light} we rephrase the causality violation as a modification of the speed of light due to short-distance non-localities and show how this effect is generic in AdS spacetimes. This observation motivates a reassessment of the causal structure on the brane which is done in \cref{sec:eft_and_subregion_entropies,sec:eft_in_double_holo} where we discuss our candidate regions.


\section{Review}
\label{sec:review}

\subsection{Karch-Randall Models}
\label{sec:KR_models}
\begin{figure}
    \centering
    
    \begin{tikzpicture}
        \begin{scope} [xshift = -4cm]
        \draw (-2,0) -- node[below]{$\mathcal M$}(0.5,0);
        \fill (0.5,-0.125) rectangle (0.75, 0.125);
        \node at (1,-1) {(a)};
        \end{scope}

        \begin{scope} [xshift = 4cm]
        \draw (-2,0) -- node[below]{$\mathcal M$}(0.5,0);
        \draw[thick] (0.5,0) -- node[below right]{$\mathcal Q$}(4,2);
        \fill[black!20!white] (-2,0) -- (0.5,0) -- (4,2) -- (4,4) -- (-2,4) -- cycle;
        \draw[dotted] (0.5,0) -- (0.5,4);
        \draw (0.5, 0.8) node[below right] {$\mu_*$} arc (90:30:0.8);
        \node at (1,-1) {(b)};
        \end{scope}

        \begin{scope}[yshift=-4cm]
        \draw (-2,0) -- node[below]{$\mathcal M$}(0.5,0);
        \draw[thick] (0.5,0) -- node[below right]{$\mathcal Q$}(4,2);
        \node at (1,-1) {(c)};
        \end{scope}
    \end{tikzpicture}
    \caption{The three descriptions in double holography. (a) The boundary description consists of a BCFT on the asymptotic boundary $\mathcal M$ coupled to a large number of degrees of freedom at its boundary (black square). (b) The bulk description is given by anti-de Sitter space with asymptotic boundary $\mathcal M$ and cut off by an end-of-the-world brane $\mathcal Q$. (c) The brane description consists of a non-gravitational CFT on $\mathcal M$ which can exchange excitations with an effective CFT on $\mathcal Q$ coupled to gravity.}
    \label{fig:three_perspectives}
\end{figure}
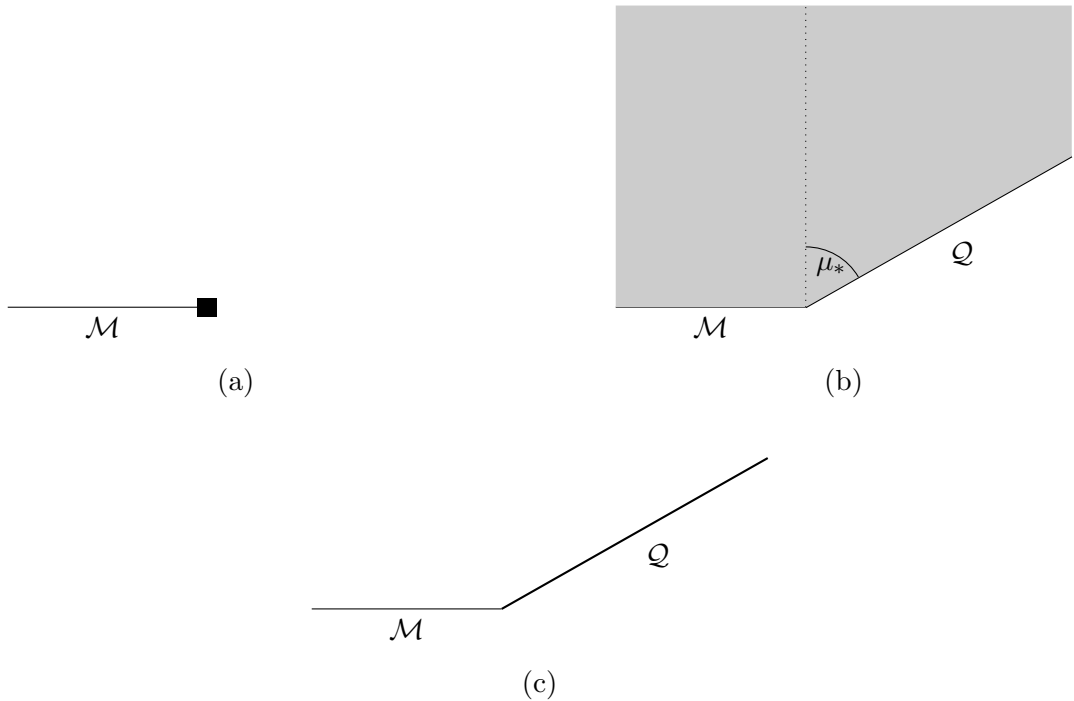
The theory under consideration in this note originates in the work by Randall and Sundrum \cite{randall1999alternative} and later Karch and Randall \cite{Karch:2000ct}, see \cref{fig:three_perspectives}. Here we will review its salient features which will be relevant for later parts of the paper.

The model consists of a $d+1$-dimensional bulk AdS spacetime with asymptotic boundary $\mathcal M$. The bulk spacetime is terminated at a finite distance by a co-dimension one hypersurface $\mathcal Q$ with constant tension $T$, called the end-of-the-world (ETW) brane. The action is given by
\begin{align}
    I  = \frac{1}{16 \pi G_\text{N}} \int d^{d+1}x \sqrt{-g}(R - 2 \Lambda) - \frac{T}{8 \pi G_\text{N}} \int_{\mathcal Q} d^dx \sqrt{-h} - \frac{1}{8 \pi G_\text{N}} \int_{\mathcal Q\cup \mathcal M} d^dx \sqrt{-h} K,
\end{align}
where $K = h^{\mu\nu} K_{\mu\nu} = h^{\mu\nu} \nabla_\mu n_\nu$ is the extrinsic curvature scalar of the (asymptotic) boundary and $G_\text{N}$ is Newton's constant in the bulk. The quantities $n_\nu$ and $h_{\mu\nu}$ denote the inwards pointing normal vector and the projector onto the boundary, respectively. In a (standard) misuse of notation we will use $h_{ij}$ for the induced metric on $\mathcal Q$. 

We choose Dirichlet boundary conditions on $\mathcal M$ and Neumann boundary conditions on $\mathcal Q$ \cite{Takayanagi:2011zk}. Upon varying the action with respect to the metric, apart from the usual bulk equation of motion, the vanishing of the boundary term at the brane implies that on $\mathcal Q$,\footnote{See for example appendix A of \cite{Lee:2022efh} for details. Note that our convention is slightly different as here we take $n$ to be inwards pointing.}
\begin{align}
    K_{\mu\nu} = - \frac{T}{(d-1)}  h_{\mu\nu}.
\end{align}

We will be interested in solutions for which the bulk metric can be written as a warped product of $AdS_d$ and an interval $\mu \in (-\frac \pi 2, \mu_*]$,
\begin{align}
\label{eqn:slicingAdS}
    ds^2 = \frac{L^2}{\cos^2 \mu} \left(d\mu^2 + ds^2_{\text{AdS}_d}\right).
\end{align}
The hypersurface $\mathcal Q$ intersects the asymptotic boundary, i.e., $\partial \mathcal Q = \partial \mathcal M$. In these coordinates $\mathcal Q$ is located at constant $\mu_*$ with $T = \sin(\mu_*) \cdot (d-1)/L  $. Since 
$\sin\mu_*<1$ for $\mu_* < \frac \pi 2$, there is a critical value $T_\text{crit} = (d-1)/L$ such that $T < T_\text{crit}$. 
 More generally, other values of $T$ are possible and correspond to branes which intersect the boundary at null- or spacelike loci. We will however not discuss those solutions here. Also note the AdS length scale for the brane (at $\mu=\mu_*$) is given by $L_\text{brane}=L/\cos\mu_*$.

As it stands, this system serves as a generic bottom-up model for holographic duals of BCFTs, which captures several important (universal) effects  \cite{Takayanagi:2011zk, Fujita:2011fp, Sully:2020pza}.
For example, the value of the tension parameter $T$ is related to a boundary central charge (or g-function) \cite{Affleck:1991tk}. Similar to usual AdS/CFT, the dual BCFT can be thought of as living on the asymptotic boundary $\mathcal M$. The intersection of the brane $\mathcal Q$ with the asymptotic boundary $\mathcal M$ defines the location of the boundary of the BCFT. In the following, we will call the gravitational description in $d+1$ dimensions the \emph{bulk description} while the BCFT formulation will be called \emph{boundary description}.

In addition to the usual duality, as $T \to T_\text{crit}$, the above model gives rise to another emergent, effective description which we will call the \emph{brane} or \emph{intermediate description}. As $T \to T_\text{crit}$, the lowest lying graviton mode localizes near the brane and plays the role of a (massive) graviton on the brane \cite{Karch:2000ct, Porrati:2001db, Aharony:2003qf, Neuenfeld:2021wbl}. It can be argued that the brane's dynamics is controlled by Einstein gravity plus higher curvature corrections which are small in the same limit \cite{deHaro:2000wj}. Thus, integrating out the region between the asymptotic boundary and the brane leaves one with a theory that consists of the boundary CFT on $\mathcal M$ glued to an effective CFT on $\mathcal Q$ coupled to gravity.\footnote{In two dimensions the gravity theory on the brane is given by Polyakov gravity \cite{Polyakov:1981rd, Chen:2020uac}. To find interesting gravitational behaviour in this case, JT gravity is usually added to the brane \cite{Almheiri:2019hni}.} The gravitational coupling on the brane is given by\footnote{For $d=2$ we can define the gravitational coupling as $G_N^\text{brane} = 2 G_N / L$ as explained in \cite{Chen:2020uac}.} $G_N^\text{brane} = (d-2) G_N/L$.
The gravitational action on the brane can be obtained via holographic renormalization and reads
\begin{align}
    \label{eq:brane_grav_action}
    I_\text{brane} = \frac{1}{16 \pi G^\text{brane}_N} \int d^d x \sqrt{-g} \left(\mathcal R + 2 \Lambda_b  + \frac{L^2}{(d-4)(d-2)} \left(\mathcal R^{ij} \mathcal R_{ij} - \frac{d}{4(d-1)} \mathcal R^2 \right) + \dots \right).
\end{align}
Here, $\Lambda_b$ is the leading order cosmological constant on the brane. The ellipsis signals higher curvature corrections. As can be seen from this action, for curvature scales of order $\mathcal O(L)$ we should expect the higher order derivative terms to become important and the description in terms of Einstein gravity together with subleading higher curvature corrections to break down. On the other hand, as long as the curvature scales are smaller than $\mathcal O(L)$, this theory can be made sense of in the usual sense as an EFT \cite{Burgess:2003jk}, where corrections to the leading order term are expanded in $E/E_\text{cutoff}$. Here, $E$ is a characteristic energy scale of the process or quantity under consideration and $E_\text{cutoff}$ is the scale at which the low energy EFT becomes susceptible to the details of its high-energy parent theory.

\subsection{The Scale of the EFT Cutoff}
\label{sec:scale_of_cutoff}
As just seen, an important aspect of doubly-holographic models is that the brane description comes with a natural UV cutoff.
We now review and extend arguments from various angles to demonstrate that the brane perspective has a cutoff scale which is set by the bulk AdS length $L$ \cite{Chen:2020uac}.

\paragraph{Bulk description}
Holographic theories in asymptotically AdS space realize the IR/UV connection \cite{Susskind:1998dq} such that an IR cutoff in the bulk corresponds to a UV cutoff in the dual boundary theory. As the brane $\mathcal Q$ removes part of the asymptotic boundary, it is natural to expect that the theory on the brane comes with a UV cutoff scale which is set by the brane's location which acts as a bulk IR regulator.

\begin{figure}
    \centering
    \begin{tikzpicture}
        \begin{scope}
        \draw (-2,0) -- node[below]{$\mathcal M$}(0.5,0);
        \draw[dashed] (0.5,0) -- node[below]{$\mathcal{M}_\text{imagined}$} (4,0);
        \draw[thick] (0.5,0) -- (4.05,2) node[above left]{$\mathcal Q$};
        \draw (3,0) arc (0:180:0.81);
        \node at (1.8,0.7) [above left] {$p$}; 
        \draw[->] (2.19,0) -- node [above] {$r$} ++(0.81,0);
        \end{scope}
    \end{tikzpicture}
    \caption{Covariant computation of cutoff on the brane. The radius $r$ of the smallest sphere on the imagined asymptotic boundary such that the entanglement wedge touches the point $p$ is $r = y_p \cot \mu_*$.}
    \label{fig:covariant}
\end{figure}
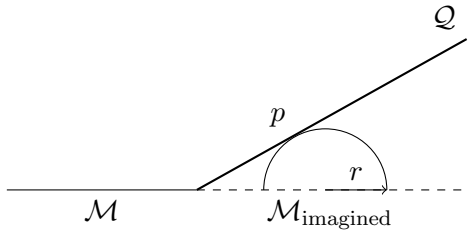

We can be more precise by utilizing the fact that in the case at hand we can extend the spacetime behind the brane and estimate the energy scale for excitations closer to the (imagined) asymptotic boundary. 
In the Poincar\'e patch, \cref{eq:poincare}, one associates a cutoff at fixed coordinate $z_c$ to a length cutoff scale of $\Lcut = z_c$. This statement can be made covariant in the following way. Given a cutoff surface in the bulk we can extend the geometry through it to an imagined boundary at $z=0$. We assign a cutoff length scale $\Lcut$ to every point $p$ on the brane which equals the size of the smallest ball-shaped region in the imagined boundary CFT whose entanglement wedge\footnote{\label{foot:EW}We remind the reader that the entanglement wedge (EW) associated with a boundary subregion $\mathcal A$ is the bulk domain of dependence of a partial bulk Cauchy slice $\Sigma$ which is bounded by $\mathcal A$ and the HRT surface associated to $\mathcal A$. We define the entanglement wedge for brane subregions analogously.} touches $p$ and is completely contained in the region between the cutoff surface and the imagined boundary. In the case of the CFT vacuum state and a constant cutoff this reproduces $\Lcut = z_c$. To find the scale on the brane with respect to the induced metric we need to Weyl transform to the metric on the brane. This changes the cutoff to $L$, the bulk AdS$_{d+1}$ length scale.

We can use the covariant formulation above to compute the cutoff on the brane of the Karch-Randall model in conformal slicing coordinates, \cref{eq:SlicingMetric}. The metric on the brane is related to the metric on the boundary, by a rescaling with a factor $\frac{L}{y \cos \mu_*}$.
Taking a boundary point $y$, the largest sphere around it has a (coordinate) radius of $y \cot\mu_*$ (\cref{fig:covariant}). We thus find that the cutoff on the brane is given by
\begin{align}
    \label{eq:cutoff_weyl}
    \Lcut^\text{brane} = \frac{L}{\sin \mu_*}.
\end{align}
Note that the cutoff diverges in the tensionless brane limit ($\mu_*\rightarrow 0$) and approaches the AdS length scale $L$ in the critical limit ($\mu_*\rightarrow \pi/2$). The divergence of the cutoff length is expected since in order to reconstruct a point on a brane with $\mu_* = 0$, we need access to all of the imagined asymptotic boundary behind the brane.

\paragraph{Brane description}
Another argument from the brane point of view was already hinted at in \cref{sec:KR_models}. In the Karch-Randall-Sundrum constructions, the gravitational action on the brane takes the form of counterterms (with the wrong sign), which can be computed using holographic renormalization \cite{deHaro:2000vlm}. The action is that of gravity plus higher curvature corrections, \cref{eq:brane_grav_action}, which are suppressed relative to the leading order Einstein-Hilbert term by
\begin{align}
    \label{eq:time_resolve}
    \epsilon \sim \left(\frac \pi 2 - \mu_* \right)^2 \sim \left(\frac {L}{L_\text{brane}}\right)^2.
\end{align}
The higher curvature corrections are of the order of $(\frac{L_\text{brane}}{\lambda})^2$, where $\lambda$ is some characteristic curvature scale of the geometry. For the higher order curvature corrections to be small, we then need that
\begin{align}
    \lambda \ll L \sim \Lcut.
\end{align}
Note that this argument relies on the brane having close-to-critical tension, i.e., $\mu_* \to \frac \pi 2$. In this limit $\Lcut$ agrees with \cref{eq:cutoff_weyl}.

Similarly, considering the graviton propagator which obtains corrections from matter loops. The leading order corrections from the insertion of a two-point function of the stress-energy tensor take the form
\begin{align}
\braket{h(p)h(-p)} \sim p^{-2} \left(1 + c G^\text{brane}_N p^{d-2} + ...\right) \sim p^{-2} \left(1 + (L \; p)^{d-2} + \dots \right).
\end{align}
As can be seen from this expression, higher order corrections become important when $ p \sim L^{-1}$. 

Note that naively, this might seem to indicate that fluctuations of the geometry become so large that the notion of geometry itself below the cutoff scale breaks down. In the examples considered in this paper, however, we are helped by the bulk description. In this paper we will be working at leading order in the $1/N$ expansion and from the bulk point of view, the geometry --- including the induced metric on the brane --- is classical and we can in fact talk about distances which are smaller than the cutoff in the brane description. However, what our discussion made clear is that we cannot expect a local theory at this scale.

\subsection{Apparent Violations of Causality on the Brane}

Gao and Wald proved that in AdS signals travel fastest along the asymptotic boundary \cite{Gao:2000ga}. This ensures that signals cannot take a shortcut through the bulk, thereby guaranteeing that holography in asymptotically AdS spaces is compatible with a dual description as a local quantum field theory. One of the crucial assumptions of the Gao-Wald theorem is that the signalling happens between points on the 
boundary located at \emph{asymptotic} infinity which fails in double holography and more generally in other models of cutoff holography.

As pointed out by \cite{Omiya:2021olc}, in double holography we can easily find examples where communication through the bulk is faster than sending signals along the boundary which questions the validity of double holography as an (approximate) local quantum field theory. 

To see how the apparent violation of causality arises, consider Poincar\'e AdS in slicing coordinates
\begin{align}
\label{eq:SlicingMetric}
ds^2 = \frac{L^2}{\cos^2 \mu} \left(d\mu^2 + \frac{-dt^2 + dy^2 + d\vec \xi^2}{y^2}\right),
\end{align}
where again $\mu \in (-\frac \pi 2, \mu_*]$ and the brane $\mathcal Q$ located at $\mu = \mu_*$ cuts off the geometry. The metric on the brane is
\begin{align}
    ds_\text{brane}^2 = \frac{L^2}{\cos^2 \mu_*}\left( \frac{-dt^2 + dy^2 + d\vec \xi^2}{y^2}\right).
\end{align}
The remaining asymptotic boundary $\mathcal M$ consisting of the half space $\mu \to - \frac \pi 2$, has the metric (after the appropriate Weyl transform)
\begin{align}
\label{eqn:boundary_metric}
    ds_\text{boundary}^2 = {-dt^2 + dy^2 + d\vec \xi^2}.
\end{align}
We will choose conventions for $y$ such that points at the asymptotic boundary are located at negative values of $y$ while points on the brane sit at $y>0$.
The metric \cref{eq:SlicingMetric} can be obtained from the usual Poincar\'e metric
 \begin{equation}
 \label{eq:poincare}
     ds^2=L^2\left(\frac{-dt^2+dz^2+d\vec{x}^2}{z^2}\right).
 \end{equation}
by the coordinate transformation
\begin{equation}
\label{eqn:Slicing_to_poincare}
     z=y\cos\mu;\quad x^1=y\sin\mu; \quad \xi^i=x^i~\text{for} ~ i=2,...,d-1.
 \end{equation}

We can now compare the time it takes for a light-like signal to reach a point $p$ on the brane from a point $q$ on the asymptotic boundary. First, let us assume that the signal travels along the asymptotic boundary to the defect and then continues on the brane. With the speed of light set to $1$, from $q$, it takes $\Delta t_1 = (y_q^2 + (\vec \xi_q - \vec \xi_0)^2)^{1/2}$ to reach the interface $y = 0$ at some $\vec \xi = \vec \xi_0$. From there it takes $\Delta t_2 = (y_p^2 + (\vec \xi_p - \vec \xi_0)^2)^{1/2}$ to reach point $p$ on the brane. Thus the minimal coordinate signaling time in the \emph{brane description} is
\begin{align}
\label{eq:dtbrane}
    \Delta t_\text{brane} = \sqrt{(y_p + y_q)^2 + (\vec \xi_p - \vec \xi_q)^2}.
\end{align}
Next, we are interested in the time it takes a light ray to propagate between $p$ and $q$ through the bulk. Realizing that the Poincar\'e AdS metric is related to the upper half-space of $\mathbb R^{d,1}$ by a Weyl transformation reduces this problem to a simple trigonometric problem in flat space, namely finding the smallest distance between $p$ and $q$ in $\mathbb R^{d,1}$. 
The coordinate signaling time equals the coordinate distance which is given by
\begin{align}
\label{eq:dtbulk}
    \begin{split}
    \Delta t_\text{bulk} &{}= \sqrt{(y_q + y_p \sin \mu_* )^2 + (y_p \cos \mu_*)^2 + (\vec \xi_p - \vec \xi_q)^2} \\
    &{}= \sqrt{(y_p + y_q)^2 - 2 y_p y_q (1 - \sin \mu_*) + (\vec \xi_p - \vec \xi_q)^2}.
    \end{split}
\end{align}
From those expressions it is easy to see that in general $\Delta t_\text{bulk} - \Delta t_\text{brane} \leq 0$ such that the signal through the bulk arrives earlier than the signal along the brane and the boundary, 
\begin{align}
\label{eq:timeadvance}
\Delta t_\text{brane}^2 - \Delta t_\text{bulk}^2 = 2 y_p y_q (1 - \sin \mu_*),    
\end{align}
leading to an apparent violation of causality in the boundary-brane system. It can also be checked that the problem of faster-than-light signalling also affects other observables, e.g., commutators \cite{Omiya:2021olc}.

\section{The Scale of Causality Violation}
\label{sec:scale_of_violation}
The presence of faster-than-light communication is problematic, since it questions the treatment of the brane theory as a quantum field theory coupled to gravity. However, in light of the discussion in \cref{sec:scale_of_cutoff} we can potentially avoid the conclusion that the brane theory is invalid, since we should not expect locality to hold below the cutoff scale of the brane EFT. And in fact, as we will now argue, the time-advance \cref{eq:timeadvance} cannot be resolved within the regime of EFT validity. This might seem surprising given that in Minkowski space deviations from the speed of light are easily detected by waiting for a sufficiently long time. However, as we will see in \cref{sec:speed_of_light}, this intuition does not apply in AdS space. In this section, we give a short calculation to demonstrate that the superluminal signalling in double holography discussed above only leads to a signal advance below the cutoff length $L/\sin\mu_*$.

\subsection{A Finite Size for Excitations}
\begin{figure}
    \centering
    \begin{subfigure}[h]{0.48\textwidth}
\hspace{0.3cm}
    \begin{tikzpicture}
        \fill[gray, opacity=0.55] (0,0) rectangle(-2,4);
        \node[below] at (-1,4.6) {boundary};
        \node[below] at (1,4.6) {brane};
        \draw (0,0) -- (-2,0);
        \fill[gray, opacity=0.3] (0,0) rectangle(2,4);
        \draw[->] (0,0) -- (0,4.5) node [above] {$t$};
        \draw[->] (0,0) -- (2.5,0) node [right] {$y$};
        \draw[dotted] (-1,0) node [below] {$y_q$} -- (-1,1);
        \fill (-1,1) circle (1mm) node[above] {$q$};
        \draw[dotted] (0.8,0) node [below] {$y_{p'}$} -- (0.8,2.8);
        \fill (0.8,2.8) circle (1mm) node[above] {$p'$};
        \draw[dotted] (1.3,0) node [below] {$y_p$} -- (1.3,2.8);
        \draw (-1,1) -- (0.8,2.8);
        \draw (-1,1) -- (1.3,2.8);
        \fill (1.3,2.8) circle (1mm) node[above] {$p$};
        \draw[dotted] (-2,1) -- (-1,1);
        \draw[dotted] (-2,2.8) -- (1.3,2.8);
        \draw [decorate,decoration = {brace}] (-2,1) -- node[left]{$\Delta t$} (-2,2.8);
    \end{tikzpicture}
    \caption{\label{fig:scale_of_causality_A}}
    \end{subfigure}
\begin{subfigure}[h]{0.48\textwidth}
\hspace{0.3cm}
    \begin{tikzpicture}
        \fill[gray, opacity=0.5] (0,0) rectangle(-2,4);
        \node[below] at (-1,4.6) {boundary};
        \node[below] at (1,4.6) {brane};
        \draw (0,0) -- (-2,0);
        \fill[gray, opacity=0.3] (0,0) rectangle(2,4);
        \draw[->] (0,0) -- (0,4.5) node [above] {$t$};
        \draw[->] (0,0) -- (2.5,0) node [right] {$y$};
        \draw[dotted] (-1,0) node [below] {$y_q$} -- (-1,1);
        \fill (-1,1) circle (1mm) node[above] {$q$};
        \fill (1.3,2.8) circle (1mm) node[below] {$p'$};
        \draw[dotted] (1.3,0) node [below] {$y_p$} -- (1.3,3.3);
        \draw (-1,1) -- (1.3,3.3);
        \draw (-1,1) -- (1.3,2.8);
        \fill (1.3,3.3) circle (1mm) node[above] {$p$};
        \draw[dotted] (-2,1) -- (2,1);
        \draw[dotted] (-2,2.8) -- (1.3,2.8);
        \draw[dotted] (1.3,3.3) -- (2,3.3);
        \draw [decorate,decoration = {brace}] (-2,1) -- node[left]{$\Delta t_\text{bulk}$} (-2,2.8);
        \draw [decorate,decoration = {brace}] (2,3.3) -- node[right]{$\Delta t_\text{brane}$} (2,1);
    \end{tikzpicture}
    \caption{\label{fig:scale_of_causality_B}}
    \end{subfigure}
    \caption{The configurations of points $q,p,p'$ used to show that causality violations by faster-than-light signalling cannot be resolved in the effective theory, since the spatial (a) and temporal (b) resolution is too low.}
    \label{fig:scale_of_causality}
\end{figure}
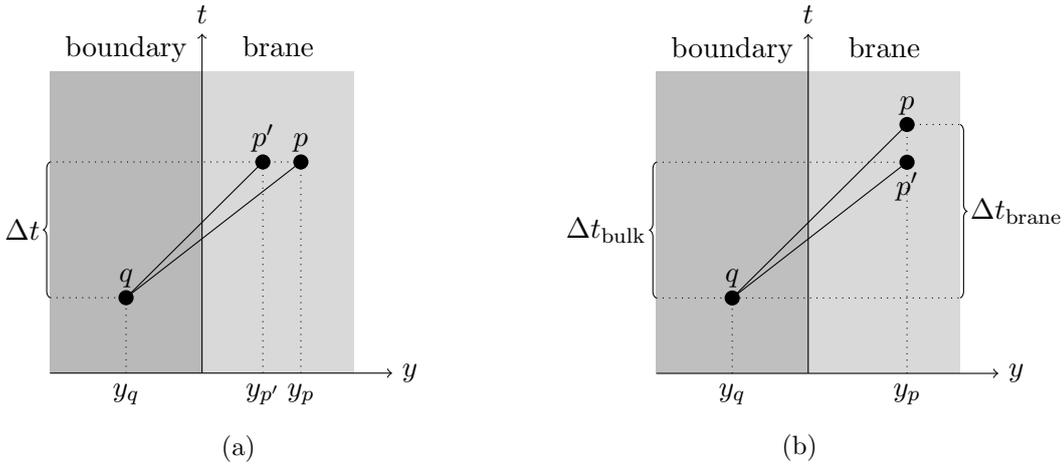

Consider a signal being sent between two points $p$ and $q$, with $q \in \mathcal M$ and  $p \in \mathcal Q$, \cref{fig:scale_of_causality_A}. In the following we will use $y_p>0$, etc., to denote the coordinates of the point $p$ in the brane description. Points in $\mathcal{M}$, e.g., $q$, will have $y=-y_q$. The signalling could be modeled by computing a two-point correlation function or a commutator between a brane and boundary operator. Due to the presence of a cutoff we are not able to answer questions which depend on the location of the point $p$ within the brane description to a precision greater than $\Lcut=L/\sin\mu_*$. The question then is how large our uncertainty about the point $p$ must be to be insensitive to the time-advance, \cref{eq:timeadvance}. To model this we will give excitations or operator insertions a finite size $s$.

We can place an upper bound on the required size $s$ of the excitation by asking about the location $y_{p'}$ at which an excitation which started at $q$ will arrive after time $\Delta t$ while propagating along the brane and boundary. The distance between $y_{p'}$ and $y_p$ is then the smallest size over which we need to smear an excitation located at $y_p$ such that we cannot tell that the signal through the bulk has arrived earlier. The location $y_{p'}$ is found by equating \cref{eq:dtbrane} and \cref{eq:dtbulk}, in which we replace $y_p$ by $y_{p'}$, and gives
\begin{align}
    y_{p'} = \sqrt{(y_p + y_q)^2 - 2 y_p y_q (1 - \sin \mu_*)} - y_q.
\end{align}
It is clear from this expression that $y_p - y_{p'}$ depends on both the location of $p$ on the brane and $q$ on the boundary. To place an upper bound, we look for a maximum of $y_p - y_{p'}$ with respect to variation of the point on the asymptotic boundary $y_q$. There is no local minimum and the global minimum is obtained by sending $y_q \to \infty$,
\begin{align}
\label{eqn:cutoffCoordSize}
    \left(y_p-y_{p'}\right)_{\text{max}}=y_p(1-\sin\mu_*).
\end{align}
This result might seem worrying at first, since the distance between the points $y_p$ and $y_p'$ depends on $y_p$ itself which is unbounded from above. However, rephrasing this expression in terms of the proper size on the brane,
\begin{align}
\label{eq:ex_proper_size}
    s/2 = \frac{L}{\cos \mu_*} \int_{p'}^p \frac{dy}{y} = - \frac{L}{\cos \mu_*} \log \sin \mu_* ,
\end{align}
we see that in fact the scale associated with the advance of the particle on the brane is coordinate independent. We also observe that $s<L/\sin\mu_*$ for $\mu_*\in (0,\pi/2)$, i.e., the proper distance is smaller than the cutoff scale \cref{eq:cutoff_weyl} on the brane. If the brane tension increases towards its maximal value, $\mu_* \to \frac \pi 2 - \delta$, we find that the proper distance is
\begin{align}
\label{eq:excitation_less_cutoff}
    s/2 \to \frac{L}{\delta} \frac 1 2 \delta ^2 \qquad \Rightarrow \qquad s \to L_\text{brane} \delta^2.
\end{align}
Thus, if we cannot localize particles within a spatial region of size $\mathcal O(L_\text{brane} \delta^2) = \mathcal O(L \delta)$, we are not able to detect a faster-than-light propagation of signals. In the near critical limit, this length scale is not only consistent with our expectation of a cutoff length of $L$, but in fact smaller by a factor of $\delta$.

\subsection{Time Advance}
Alternatively, we might ask what is the maximal proper time advance of a signal arriving from a point on the boundary to any point on the brane. We now show that the time advance that is required to accommodate causality is again smaller than the cutoff scale.

Consider \cref{fig:scale_of_causality_B}. Let point $p'$ be at the same spatial coordinate location as $p$ but with a time advance of $\Delta t_{pp'}$,
\begin{align}
\begin{split}
    \Delta t_{pp'}&=\Delta t_{\text{brane}}-\Delta t_{\text{bulk}}\\
    &= \sqrt{(y_p + y_q)^2 + (\vec \xi_p - \vec \xi_q)^2}- \sqrt{(y_p + y_q)^2 - 2 y_p y_q (1 - \sin \mu) + (\vec \xi_p - \vec \xi_q)^2}.
\end{split}
\end{align}
This is the difference in coordinate time with which a detector at $y_p$ would detect signals which propagate according to boundary and bulk causality, respectively.
For any $(\vec \xi_p - \vec \xi_q)$ separation, $\Delta t_{pp'}$ is a monotonically increasing function of $y_q$ which is maximized as $y_q\rightarrow\infty$ to give an upper bound
\begin{align}
    \Delta t_{pp'}^{\text{max}}=y_p(1-\sin\mu_*).
\end{align}
Note that this is exactly the same coordinate size as in \cref{eqn:cutoffCoordSize}. Once again, it is instructive to compute the proper time advance $\tau$ which highlights that it depends solely on the location of the brane and is independent of the spatial position of the point $p$ itself on the brane. Using the above result with the metric in \cref{eq:SlicingMetric}, we obtain
\begin{align}
\label{eqn:TimeAdvance_proper}
    \tau=L \int_{0}^{\Delta t_{pp'}^{\text{max}}}\frac{dt}{y_p \cos\mu_*}=L\left(\frac{1-\sin\mu_*}{\cos\mu_*}\right).
\end{align}
We can once again observe that $\tau\,(\text{or }2\tau)<L/\sin\mu_*$ for $\mu_*\in (0,\pi/2)$, i.e., the required time advance is smaller than the cutoff scale \cref{eq:cutoff_weyl} on the brane. If the brane tension increases towards its maximal value, $\mu \to \frac \pi 2 - \delta$, we find that the required proper time advance is
\begin{align}
    \tau\sim L\frac{\delta}{2}\sim L_{\text{brane}}\frac{\delta^2}{2}.
\end{align}
Hence the required time advance is smaller than the cutoff by a factor of $\delta$ just as in \cref{eq:excitation_less_cutoff}. In summary, we conclude that although there seemingly is faster-than-light signalling in the brane description, it cannot be resolved within the regime of validity of the brane theory.

\section{AdS, Cutoffs and the Causal Structure}
\label{sec:speed_of_light}
The discussion in \cref{sec:scale_of_violation} makes it clear that the violation of causality found in \cite{Omiya:2021olc} is outside the regime of validity of the effective theory in the brane perspective. This, however, still leaves a puzzle.

Effective field theories can still be written as local field theories despite the existence of a UV cutoff. This is precisely the standard Wilsonian approach to QFT. Although the theory might break down at small length scales (and we do not necessarily expect locality to hold at scales below the cutoff), we can still talk about the causal structure induced by the background on which the QFT is placed. 

Since we are working at leading order in $1/N$, the bulk geometry and thus also the brane geometry are frozen. 
Therefore, despite the fact that any experiment that resolves the time-advance of signals cannot be described within the brane effective field theory, subregions and light cones can still be given a well defined notion, even below the cutoff scale. For example, given some initial data on an achronal subregion $A$, we can ask about its future, i.e., the region in which the solution is affected by the choice of initial data on $A$. 

In flat space, one is used to the fact that the speed of light in the low energy EFT agrees with the speed of light in the parent theory which is valid at higher energies and so the causal structure is still plausibly given by the background metric, at least as long as we consider large distances. However, as we will now discuss, in double holography the future of a region $A$ can be larger than naively thought since the effective speed of signal propagation on the brane can exceed the bulk/boundary speed of light $c$. Since the future of $A$ is larger than thought, the region in which we can uniquely predict the solution given initial data on $\bar A$, the complement of $A$, is smaller than naively assumed. We thus need to conclude that given some initial data on a subregion $A$, the spacetime region in which we can uniquely predict the physics is not bounded by light cones with the usual speed of light. In other words, the causal structure on the brane needs to be modified. Before we formulate this problem more precisely in \cref{sec:eft_and_subregion_entropies,sec:eft_in_double_holo}, we will demonstrate here that the speed of signal propagation on the brane is faster than $c$, however, only in one direction. We will then use a simple toy model to argue that such an effect is natural for EFTs placed on an AdS background.

\subsection{Faster-than-light Light Fronts}
\label{sec:a_puzzle}

\begin{figure}
    \centering
        \begin{subfigure}[h]{0.47\textwidth}
\hspace{0.3cm}
    \begin{tikzpicture}
        \fill[left color=gray!10!white, right color = gray!70!white] (-2,0) -- (0.5,0) -- (2.25,1) -- (2.25,4) -- (-2,4) -- cycle;
        \draw (2.25,1) -- (2.25,4);
        \draw [->] (2.5, 2.5) -- (2.75,2.5) node[right] {$c$} ;
        \draw [->] (2.5, 3.5) -- (2.75,3.5);
        \draw [->] (2.5, 1.5) -- (2.75,1.5);
        \draw (0,0) -- (-2,0);
        \draw (0,0) -- (0.5,0);
        \draw[thick] (0.5,0) -- (4,2);
        \draw[->] (1,0) -- node[below right] {$c^R_\text{eff} > c$} (1.875,0.5);
    \end{tikzpicture}
    \caption{\label{fig:left_to_right}}
    \end{subfigure}
        \begin{subfigure}[h]{0.47\textwidth}

    \begin{tikzpicture}
        \fill[right color=gray!10!white, left color = gray!70!white] (4,4) -- (1,4) -- (1,2) to [out = -90, in=120, looseness=1] (1.4,0.5) -- (4,2) -- cycle;
        \draw [<-] (0.5, 2.5) node[left] {$c$} -- (0.75,2.5);
        \draw [<-] (0.5, 3.5) -- (0.75,3.5);
        \draw [<-] (0.5, 1.5) -- (0.75,1.5);
        \draw [<-] (0.7, 0.5) -- (0.95,0.64);
        \draw (0,0) -- (-2,0);
        \draw (0,0) -- (0.5,0);
        \draw[thick] (0.5,0) -- (4,2);
        \draw[<-] (1,0) -- node[below right] {$c^L_\text{eff} = c$} (1.875,0.5);
    \end{tikzpicture}
    \caption{\label{fig:right_to_left}}
    \end{subfigure}
    
    \caption{We can send a plane wave to the left or to the right. 
    (a) If we send the light front from the boundary towards the brane, the light front on the brane moves at $c_\text{eff}$. 
    (b) If we send it from the brane towards the asymptotic boundary, the light front moves with $c$.}
    \label{fig:enter-label}
\end{figure}
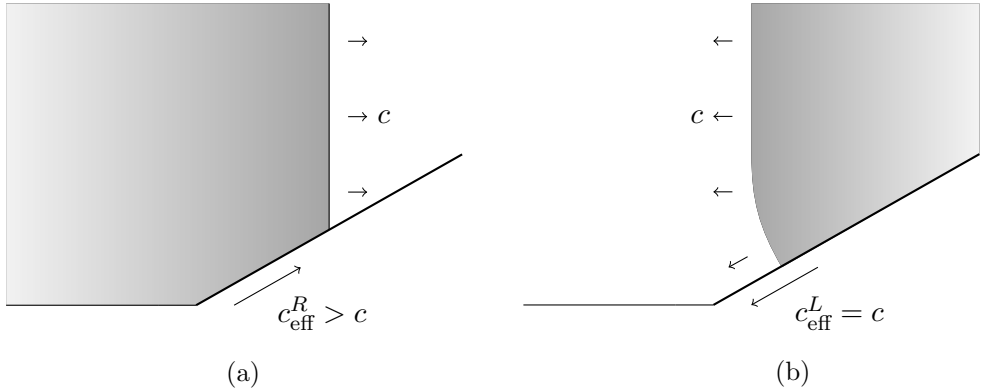
An easy way to see that the causal structure on the brane is modified is by reconstructing bulk operators in the entanglement wedge that corresponds to the region $y\in(-\infty, -y_{q})$ on the asymptotic boundary. In the brane description the correct entanglement wedge does not connect to the brane \cite{Neuenfeld:2021bsb} and we can create any state in the bulk which has support on $x = y \sin \mu  < -y_{q}$. In particular this will allow us to create a plane wave travelling parallel to the asymptotic boundary towards the brane. As the plane wave passes $x = 0$ it starts colliding with the brane. The intersection of this plane wave with the brane moves at an effective speed $c^R_\text{eff}$ (\cref{fig:left_to_right}). One can easily compute the intersection of the plane wave with the brane and finds 
\begin{align}
    \label{eq:effective_c}
    c^R_\text{eff} = \frac{c}{\sin \mu_*}.
\end{align}
In the near-critical limit of the brane tension, this implies that $c^R_\text{eff} \to 1 + \frac {\delta^2} 2 + \mathcal O(\delta^4)$, where we remind the reader that $\delta = \frac \pi 2 - \mu_*$.

We can run the same argument by starting with a region on the brane and sending a plane wave (or really any signal) towards the boundary. If we assume EW reconstruction on the brane, we can use access to a brane subregion $y \in (y_p, \infty)$ to create a plane wave which travels to the left (\cref{fig:right_to_left}) and sits at $x > y_p \sin \mu_*$. Such a plane wave does not collide with the brane as before. Instead, due to what essentially amounts to Huygens principle, a new wave-front is created at the intersection of the EW with the brane  which travels along the brane with the bulk speed of light $c$.\footnote{The intersection of the right going wave and the brane also sends out wave-fronts with the bulk speed of light (Huygens principle), but similar wave-fronts are sent out by other points on the wave that collide with the brane at an angle making the effective speed $c/\sin\mu_*$. There is no modification (from $c$) to the propagation speed for the left going wave since there is no such collision with the brane. } 
We therefore must conclude that on the brane, the future of a region towards smaller values of $y$ is lower bounded by the bulk speed of light $c_\text{eff}^L = c$, while the future of a region to larger values of $y$ is bounded by  $c^R_\text{eff} > c$. This suggests that the causal structure of the theory is modified with respect to the light cones obtained from the background metric and moreover in an asymmetric way. In the following we define the effective speed of light on the brane as  
\begin{equation}
\label{eqn:ceff}
    c_\text{eff}=
    \begin{cases}
        c_\text{eff}^{R} & \text{for signals propagating away from the defect}\\
        c_\text{eff}^{L} & \text{for signals propagating towards the defect.}
    \end{cases}
\end{equation}

In the context of double holography, one might be tempted to suggest that the faster speed of light should be interpreted as a quantum gravitational effect \cite{Omiya:2021olc}. However, as we will see, also in the present context, this is \textbf{not} a (quantum) gravitational effect, but simply an implication of the AdS background geometry.

\subsection{A Simple Model for Cutoff Theories on AdS}
\label{sec:simple_model}

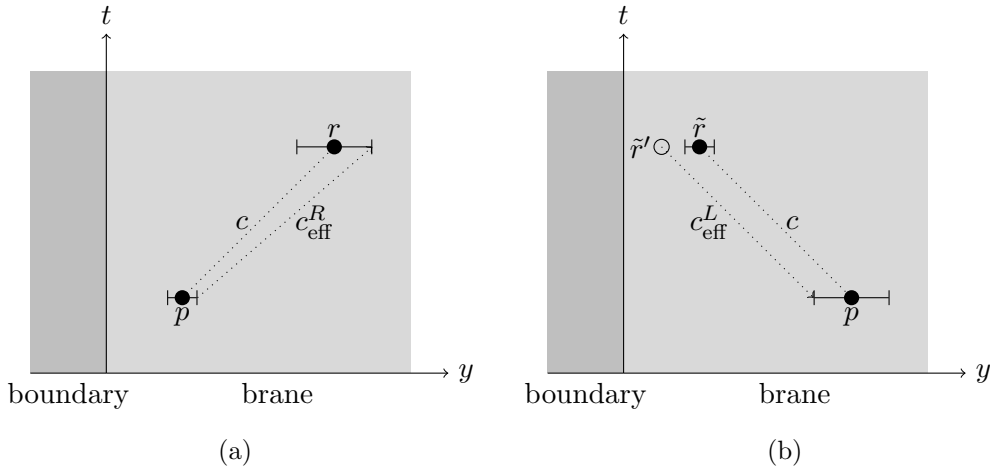
\begin{figure}
    \centering
\begin{subfigure}[h]{0.47\textwidth}
\hspace{0.3cm}
    \begin{tikzpicture}
        \fill[gray, opacity=0.5] (0,0) rectangle(-1,4);
        \draw (0,0) -- node [below] {boundary} (-1,0);
        \fill[gray, opacity=0.3] (0,0) rectangle(4,4);
        \draw[->] (0,0) -- (0,4.5) node [above] {$t$};
        \draw[->] (0,0) -- node[below] {brane} (4.5,0) node [right] {$y$};
        \fill (1,1) circle (1mm) node [below] {$p$};
        \draw[|-|] (0.8,1) -- ++(0.4,0);
        \fill (3,3) circle (1mm) node [above] {$r$};
        \draw[|-|] (2.5,3) -- ++(1,0);
        \draw[dotted, ->] (1,1) -- node[left] {$c$} (3,3);
        \draw[dotted, ->] (1.2,1) -- node[right] {$c^R_\text{eff}$} (3.5,3);
    \end{tikzpicture}
    \caption{\label{fig:model}}
        \end{subfigure}
        \begin{subfigure}[h]{0.47\textwidth}
    \begin{tikzpicture}
        \fill[gray, opacity=0.5] (0,0) rectangle(-1,4);
        \draw (0,0) -- node [below] {boundary} (-1,0);
        \fill[gray, opacity=0.3] (0,0) rectangle(4,4);
        \draw[->] (0,0) -- (0,4.5) node [above] {$t$};
        \draw[->] (0,0) -- node[below] {brane} (4.5,0) node [right] {$y$};
        \fill (1,3) circle (1mm) node [above] {$\tilde r$};
        \draw[|-|] (0.8,3) -- ++(0.4,0);
        \fill (3,1) circle (1mm) node [below] {$p$};
        \draw[|-|] (2.5,1) -- ++(1,0);
        \draw[dotted, ->] (1,3) -- node[right] {$c$} (3,1);
        \draw[dotted, ->] (0.5,3) -- node [left] {$c_\text{eff}^L$} (2.5,1);
        \draw (0.5,3) circle (1mm) node [left] {$\tilde r'$};
    \end{tikzpicture}
    \caption{\label{fig:model_right_to_left}}
\end{subfigure}
    \caption{A simple model for finite size excitations on a AdS Karch-Randall brane. The center of excitations, denoted by the black dot, moves at bulk the speed of light $c$. The leading edge moves with $c_\text{eff}$.}
\end{figure}
We can use a simple toy model to explain that theories with a minimal length scale, when placed on AdS, have an effective speed of light that naively can exceed the bulk speed of light $c$ for signalling in one direction. Consider the situation shown in \cref{fig:model} where an excitation is sent towards larger values of $y$. The excitation starts out at $p$ and has a constant proper size. Here, we decided to start the excitation on the brane, but we might as well have started on the asymptotic boundary. In our model we assume that all signals move at constant speed $c$. This holds in particular for every point if the region around $p$ that makes up the excitation. However, due to its proper size given by \cref{eq:ex_proper_size} the coordinate size of the excitation grows while moving.\footnote{Note that it is not necessary that the size of the excitation is the maximal size, \cref{eq:ex_proper_size}. In fact, as we have seen, the size is the maximal size needed to make things consistent with causality. In particular, the size might be much smaller. As found by \cite{Omiya:2021olc}, naive signal propagation between points on the brane happens with $c$, whereas violation of causality only happens when we send signals between the brane and the asymptotic boundary.} As indicated in the figure, due to this effect, the leading edge of the excitation, which in our model plays the role of $c^R_\text{eff}$, propagates faster than the bulk speed of light $c$. 

The value of $c^R_\text{eff}$ follows from a simple geometric analysis and depends on the minimal size of the excitation, \cref{fig:model}. We model the excitation by the movement of a rod of fixed proper size $s$ between 2 points $p$ and $r$ in AdS.
\begin{align}
    s/2 = \int_{y_p}^{y_{p'}} dy \frac {L_\text{brane}} y = L_\text{brane} \log \frac{y_{p'}}{y_p}. 
\end{align}
In particular, if $p$ is at coordinate location $y_p$, the right end of the rod through $p$ sits at 
\begin{align}
    y_{p'} = e^{\frac s {2L_\text{brane}}} y_p,
\end{align}
and similarly for $r$. Given the coordinate locations, we can simply compute the effective speed of light,
\begin{align}
    \label{eq:ceff_from_model}
    c^R_\text{eff} = \frac{y_{r'} - y_{p'}}{\Delta t_{pr}} =  e^{\frac s {2L_\text{brane}}} \frac{y_{r} - y_{p}}{\Delta t_{pr}} = e^{\frac s {2L_\text{brane}}} \cdot c = \frac{c}{\sin \mu_*}.
\end{align}

In the last step we substituted the minimal size of excitations needed for consistency with causality, \cref{eq:ex_proper_size}, into \cref{eq:ceff_from_model}. We reproduce \cref{eq:effective_c} and conclude that excitations whose proper size is the smallest possible size required to make double holography consistent with causality imply an effective speed of light which again agrees with the observation in double holography.

The toy model also correctly captures the light front if we signal from the brane to the boundary, \cref{fig:model_right_to_left}. The coordinate size of an excitation shrinks if it travels towards the left and so one might naively think that now the effective speed of light is smaller than $c$. However, since all signals travel at $c$, placing an excitation at location $p$ on the brane still affects the solution at $\tilde r'$, as indicated in \cref{fig:model_right_to_left}. Therefore we have that $c_\text{eff}^L = c$.

\subsection{Beyond AdS: Comment on Flat Branes}
To illustrate that the change in the effective speed of light crucially relies on the AdS background geometry we can repeat the analysis for a Randall-Sundrum brane \cite{randall1999alternative}, which is simply a brane sitting at a fixed value of $z_c$ in Poincar\'e coordinates, \cref{eq:poincare}. The brane geometry in this case is flat. This theory again has a cutoff scale of the order of the bulk AdS scale $L$. Now however, repeating the bulk computation of \cref{sec:a_puzzle} one finds that the effective speed of light on the brane coincides with the bulk $c$. This can also be reproduced by applying the model \cref{sec:simple_model} in the limit where the distance between the endpoints of the signalling, $p$ and $r$, is much larger than the cutoff scale. A simple way to see that the effective speed of light on a flat brane is $c$ (unmodified) in the simple model is that a fixed proper size $s$ would imply  a fixed (Poincar\'e) coordinate size on the flat brane. Thus, on a flat brane, the coordinate size of the excitation at $p$ would be the same as the coordinate size of the excitation at $r$ in \cref{fig:model}. Hence, the effective speed of light for a flat brane still is $c_\text{eff}=c$.

\section{Effective Field Theory and Domains of Dependence}
\label{sec:eft_and_subregion_entropies}
We have seen in the previous section that the causal structure on the brane is modified. We would now like to understand what could replace the old causal structure. At leading order in $1/N$ we have a fixed background geometry. For quantum field theory in a fixed background the causal structure is usually determined by the domain of the dependence $\mathcal D(B)$ associated with a partial Cauchy slice $B$. The region $\mathcal D(B)$ is defined as the set of points for which all past or future directed, inextensible causal curves must pass through $B$. Here, the notions of \emph{causal} and \emph{domain of dependence} are defined with respect to the causal structure provided by the induced metric on the brane. The importance of the domain of dependence is that one usually operates under the assumption that the state of the physical system in $\mathcal D(B)$ is fully determined by an initial value problem on $B$ \cite{Wald:1984rg}. In our case the initial data is given by a state\footnote{Strictly speaking this statement does not make much sense in continuum quantum field theory, since there is no density matrix of subregions. However, we will simply assume that there exists a UV regulator at length scales much below the cutoff length scale $\Lcut$ such that we can treat the theory in $\mathcal D(B)$ as the long distance limit of a finite dimensional quantum system for which a density matrix exists.} $\rho_B$. 

However, the previous subsection demonstrates that this is not correct in the brane EFT. Consider a partial Cauchy slice $B$ located at constant Poincar\'e time. As argued above, signals can effectively move with a maximum speed $c_\text{eff} > c$ and thus enter $\mathcal D(B)$ without neccessarily intersecting $B$. This clearly shows that initial data on $B$ is not sufficient to fully determine the state in $\mathcal D(B)$. One may then ask about the set of partial Cauchy slices $B'$ with $\mathcal D(B) = \mathcal D(B')$ for which we can consistently evolve the state. The collection of $B'$ for which we can do this defines the largest possible regime of predictability of the brane theory starting from initial conditions on $B$.

In this section we will define and discuss three regions which are plausible candidates to replace the domain of dependence on the brane, before computing (bounds on) those regions in \cref{sec:eft_in_double_holo}.

\subsection{Local Unitarity and the Unitary Domain of Dependence $\mathcal C$}
\label{sec:local_unitarity}
Given some initial data on a partial Cauchy slice $B$, we can give a necessary condition for predictability within the EFT description, which formalizes the fact that time evolution is implemented by a unitary operator. Consider a state $\rho(0) = \rho_B$ on a partial Cauchy slice $B$ at constant time $t_0$. The state $\rho(s_0) = \rho_{B'}$ on a partial Cauchy slice $B'$ with $\mathcal D(B) = \mathcal D(B')$ can only be obtained within EFT if there is a family of unitary operators $U(s)$ with $0 \leq s \leq s_0$ such that 
\begin{align}
\label{eq:unitary_time_evo}
\rho_{B'} = U(s_0) \rho_{B} U^\dagger(s_0).
\end{align}
Additionally, for any intermediate $0 \leq s \leq s_0$ there must be an intermediate Cauchy slice $B_s$ which interpolates between $B$ and $B'$ such that $\rho_{B_s} = U(s) \rho_{B} U^\dagger(s)$. For each of the intermediate slices  \cref{eq:unitary_time_evo} must also hold. We will call this condition \emph{local unitarity}. Generally, every choice of intermediate Cauchy slices also implies a choice of intermediate unitaries $U(s)$, but $U(s_0)$ must be unique.

Violating \cref{eq:unitary_time_evo} means non-conservation of probabilities and thus a breakdown of our predictive power using an EFT description. While it is not guaranteed that we can write down a consistent EFT for the set of slices $B_s$ where \cref{eq:unitary_time_evo} is obeyed, it is at least not obviously impossible. We will not attempt to give a precise characterization of slices $B'$ for which \cref{eq:unitary_time_evo} holds. Instead, we would like to constrain the spacetime region in which \cref{eq:unitary_time_evo} can possibly be valid thereby giving an upper bound on the region in which predictability via an EFT description could be attainable. For this the following definition, which can roughly be thought of as a replacement for the domain of dependence $\mathcal D(B)$ in the brane description, will be convenient.

\paragraph{Definition 1:} Given a partial Cauchy slice $B$ at a constant Poincar\'e time $t_0$, the \emph{unitary domain of dependence} $\mathcal C(B)$ is the union of all points $p$ such that there exists a Cauchy slice $B'$ with $\mathcal D(B) = \mathcal D(B')$ passing through $p$ and the state on $B'$ is obtained from $\rho_B$ by the action of unitaries via \cref{eq:unitary_time_evo}.
\vspace{0.5em}

The choice for $B$ to be at constant Poincar\'e time is natural in the context of double holography, since the cutoff $\Lcut$ is constant in Poincar\'e coordinates. Let us moreover emphasize that we cannot necessary predict the state on every Cauchy slice contained in $\mathcal C(B)$. Our condition does mean, however, that for any given point $p \in \mathcal C(B)$ one can find a Cauchy slice $\Sigma$ that runs through $p$ and on which the above conditions holds.

It is clear from the definition that outside the unitary domain of dependence, we do not have any predictive power via a unitary EFT description. Our condition implies that a local EFT description of the physics in a subregion which is non-locally coupled to a far away region is not possible. In practice, for a local, UV complete theory such as a conformal field theory we have that $\mathcal C(B) = \mathcal D(B)$. However, as we will see in the next section, $\mathcal C(B)$ is a proper subset of $\mathcal D(B)$ in double holography. 


\subsection{The Holographic Domain of Dependence $\mathcal H$}
The unitary domain of dependence provides an alternative to the notion of domain of dependence that leverages the property of unitarity of the effective theory. Similarly we could leverage the holographic duality with the bulk to define another potential alternative to the domain of dependence on the brane. In usual AdS/CFT setups, given a boundary subregion $B$, the entanglement wedge associated to it is the bulk region that can be fully reconstructed just from the data on $B$ \cite{Czech:2012bh,Headrick:2014cta,Wall:2012uf,Jafferis:2015del,Dong:2016eik,Cotler:2017erl}. In particular, the intersection of the entanglement wedge with the boundary is the maximal boundary region which can be described with just data on $B$. Assuming that this is still true in the presence of the brane and with the subregion $B$ now lying on the brane, we are presented with another candidate domain of dependence on the brane, which we will refer to as the holographic domain of dependence, defined as follows.

\paragraph{Definition 2:} Given a partial Cauchy slice $B$ at constant Poincar\'e time, the \emph{holographic domain of dependence} $\mathcal H(B)$ is the intersection of the entanglement wedge associated to $B$ and the brane, i.e., $\mathcal{H}(B)=\mathcal{W_E}[B]\cap \mathcal{Q}$ with 
$\mathcal{Q}$ and $\mathcal{W_E}[B]$ being the brane and the entanglement wedge associated to $B$ respectively.
\vspace{0.5em}

This region was also discussed in \cite{Omiya:2021olc} in double holography, in \cite{Lewkowycz:2019xse} in connection to $T \bar T$ and \cite{Grado-White:2020wlb} in connection with general cutoff theories. The holographic domain of dependence is a natural replacement for the domain of dependence from the holographic point of view leveraging the duality between the brane and the bulk perspectives in double holography. Note that $\mathcal{H}(B)$ and $\mathcal{C}(B)$ are defined independently and, a priori, there is no natural relationship between them. Interestingly, we will see that both these candidate domains of dependence coincide when we are dealing with a $d=2$ dimensional brane.


\subsection{The EFT Domain of Dependence $\mathcal E_\alpha$}
\label{sec:exp_from_eft}
The unitary and holographic domains of dependence defined above give an upper bound on a subregion associated to a partial Cauchy slice for which making predictions using a unitary or holographic EFT which is ``local'' above some scale is possible. However, those regions do not have to be the same as the regions for which we usually trust effective field theory.

While for the purpose of calculations in momentum space, e.g., cross section, the range of validity of effective field theory is well understood, in position space there are several different proposals for when EFT breaks down, see e.g., \cite{Polchinski:1995ta, Lowe_Thorlacius, Giddings, Kiem:1995iy}. Importantly, there is no known first principles derivation of some full set of consistency conditions under which an EFT description (in particular in the presence of gravity) is valid. 

One popular condition, which we will focus on here, is the so-called \emph{nice slice criterion} \cite{Polchinski:1995ta}. It states that a good EFT description only exists on Cauchy slices for which all local curvatures are smaller than the UV cutoff of the EFT. The precise quantitative formulation of this statement is somewhat unclear and different versions exist in the literature. This condition was used in the context of the black hole information paradox \cite{Polchinski:1995ta} or to argue for corollaries of the Quantum Focusing Conjecture \cite{Bousso:2022tdb}. 

Here, we will interpret the nice slice condition as the statement that the intrinsic and extrinsic curvature scalars of a Cauchy slice have to be much smaller than the cutoff of the theory at every point for it to be a nice slice, i.e.,
\begin{align}
\label{eq:criterion}
|\mathfrak R(\vec w)|, |\mathfrak K(\vec w)| \ll \frac 1 {\Lcut},
\end{align}
where $\mathfrak{R}(\vec w)$ and $\mathfrak{K}(\vec w)$ are the intrinsic and extrinsic curvature scalars of the Cauchy slice, respectively, and $\vec w$ are coordinates on the Cauchy slice. Of course, the choice of order $\mathcal O(1)$ factors in this expression is quite arbitrary. It would additionally be possible that further powers of $\Lcut/L_\text{brane}$ appear. However, we will give a motivation for their absence below. This condition then suggests the following definition for a spacetime region, within which we would like to trust time evolution within some EFT.

\paragraph{Definition 3:} Consider a partial Cauchy slice $B$ located at constant Poincar\'e time. The \emph{EFT domain of dependence} $\mathcal E_\alpha(B)$ is the union of all points $p$ such that there exists a Cauchy slice $B'$ with $\mathcal D(B) = \mathcal D(B')$ passing through $p$ and 
\begin{align}
   |\mathfrak R(\vec w)|, |\mathfrak K(\vec w)| \leq \frac{\alpha}{\Lcut},
\end{align}
where $\vec w$ are coordinates on $B'$, $\alpha$ is some real number and $\Lcut$ is the cutoff scale of the theory. In other words, $B'$ is a nice slice. 
\vspace{0.7em}

\begin{figure}
    \centering
    \begin{tikzpicture}
        \draw (-3,0) node [below left] {$p_1$} -- node[below] {$B$} (3,0) node [below right] {$p_2$};
        \draw  plot [smooth] coordinates {(-3,0) (0.3,2.1)  (3,0)};
        \filldraw (0.8,2.2) circle (1pt) node[anchor=west]{$p$};
        \draw (-0.8,1.4) node [above left] {$\partial^+\mathcal{E}_{\alpha}(B)$};
    \end{tikzpicture}
    \caption{$\partial^+\mathcal{E}_{\alpha}(B)$ is the future boundary  on the EFT domain of dependence of $B$.}
    \label{fig:EFT_DoD_fig}
\end{figure}
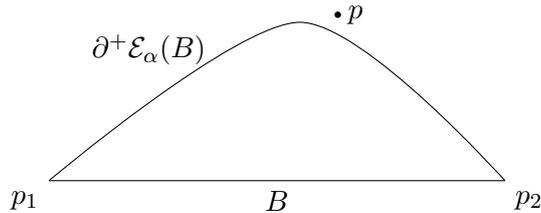

If the nice slice criterion indeed was a necessary requirement for a good EFT description, this definition would tell us that in the region $\mathcal D(B) \setminus \mathcal E_\alpha(B)$ the EFT must fail. In other words, for any point $p$ in $\mathcal{D}(B)$ that is beyond the boundary of $\mathcal{E}_{\alpha}(B)$ 
(see \cref{fig:EFT_DoD_fig}), the EFT description is expected to become unreliable when describing the dynamics starting from initial data on $B$.

Note that $\mathcal E_\alpha(B)$ depends on the choice of a parameter $\alpha$. Depending on what the value of $\alpha$ is, our condition becomes more permissive (large $\alpha$) or less permissive (small $\alpha$).
 This reflects the fact that the value for which higher curvature corrections become important and predictability of the EFT breaks down depends on the precise form of the higher curvature corrections as well as the precise question being asked. Generally speaking, we should have $\alpha \ll 1$ to be sure that EFT works.\footnote{However, depending on the question one is interested in, $\alpha$ is also chosen large sometimes \cite{Bousso:2022tdb}.}

As already mentioned previously, while it sounds plausible, a formal derivation of the nice slice criterion in the general case is not known. Moreover, we are also not aware of an argument in favour of the nice slice condition which applies to an arbitrary spacetime. In this paper, we are however not interested in the most general case. Instead we focus on the validity of EFT on subregions of vacuum AdS. In this case (as well as in the case of a flat background) we can motivate a bound on the curvatures of Cauchy slices, \cref{eq:criterion}, as follows.

First, note that via the Gauss-Codazzi equations, and the fact that we are in vacuum AdS, the intrinsic and extrinsic curvature of a Cauchy slice scale the same up to terms of the order of $L^{-1}_\text{brane} \ll L^{-1}$. It is therefore sufficient to argue that either the intrinsic or the extrinsic curvature must be smaller than $\Lcut^{-1}$.

Here we give an argument for the extrinsic curvature. Consider an EFT for a scalar field theory on an AdS background with action given by
\begin{align}
\label{eqn:scalar_field_action_on_brane}
    S \sim \int dt\; dy \;d^{d-2}\xi \sqrt{-h} \left( \phi \Box \phi - m^2 \phi^2 - \lambda \phi \Box^2 \phi + \dots \right).
\end{align}
Such an action naturally appears in Karch-Randall braneworlds and governs the dynamics of sources, see e.g., \cite{Neuenfeld:2021wbl}. We will not be interested in factors of $\mathcal O(1)$ which we will routinely set to one in the following discussion. The important piece of the above action is the term proportional to $\lambda$ which is higher order in derivatives. The coupling $\lambda \sim \mathcal O(L^2)$ is dimensionful. The ellipsis indicates that there are terms of higher order in derivatives with couplings proportional to higher powers of $L$. In the near-critical limit of the brane tension all higher derivative terms will be suppressed by $(L/\ell)^{2n}$, where $\ell$ is some characteristic length scale, such as the mass of the field $\phi$ or the AdS length $L_\text{brane}$ of the background. The above action is a good description of the theory in the sense of an EFT as long as the higher derivative terms remain small. If the term proportional to $\lambda$ becomes $\mathcal O(1)$, all other higher order terms generally also become important and the expansion of the action in terms of higher derivative corrections breaks down.

Let us now focus on this theory in a strip-shaped subregion $B$ of Poincar\'e AdS such that the holographic coordinate is bounded $y_0 \leq y \leq y_1$ while the transverse spatial coordinates $\vec \xi$ are unrestricted. We can choose a coordinate system $(\tau, u, \xi)$ which covers the domain of dependence $\mathcal D(B)$. The explicit coordinate change is given by
\begin{align} 
    \label{eq:new_coords}
    \tanh(u^\pm) = \frac{2y^\pm}{y_1-y_0},
\end{align}
where $y^\pm = t \pm (y - \frac{y_1 + y_0}{2})$ and $u^\pm = \tau \pm u$. The new coordinates $\tau$ and $u$ take values in the reals and $u = \tau = 0$ is the center of $\mathcal D(B)$. The transverse coordinates along the strip are unaffected by the coordinate transformation. In $d=2$ this new coordinate system foliates the causal diamond by constant extrinsic curvature slices where increasing $\tau$ means going to higher extrinsic curvature. For general dimensions those slices are not constant extrinsic curvature anymore. Our aim is now to understand how far we can evolve the state in $\tau$ before the higher derivative terms become dominant and relate this to a maximally admissible scale for the extrinsic curvature. 

In coordinate system \cref{eq:new_coords} the Lagrangian and therefore also the Hamiltonian which is used to evolve the state between slices are time dependent. Since the potential terms in the Hamiltonian and Lagrangian are the same (up to a sign) it is sufficient to analyze the Lagrangian.

For simplicity we will only consider the Lagrangian at late times $\tau \gg 1$ around $u = 0$. To leading order, the kinetic term takes the form
\begin{align}
    \mathcal L_\text{kin} \sim - (\partial_\tau \phi)^2 + (\partial_u \phi)^2  + \mathcal O(e^{- 2 \tau}).
\end{align}
On the other hand, the higher derivative corrections to the Lagrangian schematically take the form
\begin{align}
    \mathcal L_{\lambda}(\phi, \partial_\tau^{4}\phi, \dots ;\tau,0) \sim \frac{\lambda}{L^2_\text{brane}} e^{4 \tau} \left( \partial_\tau^{4}\phi(\tau, 0) + \dots \right),
\end{align}
where the ellipsis denotes other derivative terms with $\mathcal O(1)$ prefactors. Thus, assuming that the derivative terms are of $\mathcal O(1)$, the effective coupling for higher derivative terms is $\lambda e^{4 \tau}/ L^2_\text{brane}$ and the theory breaks down when $\lambda \sim L^2 \sim L^2_\text{brane} e^{-4\tau}$. Computing the extrinsic curvature of the slice and evaluating it around $u=0$ we find that $\mathfrak K(\tau, 0) \sim e^{2\tau}/L_\text{brane}$. This motivates the claim that $\mathfrak K \lesssim \mathcal O(1/L)$ for EFT to be valid.

\section{EFT in Double Holography}
\label{sec:eft_in_double_holo}
In this section we will investigate the general causal structure of the brane description by studying the unitary domain of dependence, the holographic domain of dependence and the EFT domain of dependence defined in the previous section. In the cases we study we will find that the usual notion of a domain of dependence $\mathcal D(B)$ disagrees with all three of these.

As we will see, all regions we discuss are smaller than $\mathcal D(B)$ and exclude those parts of $\mathcal D(B)$ which can be affected by superluminal signalling with $c_\text{eff}$. We will demonstrate this by computing $\mathcal H(B)$ and $\mathcal E_\alpha(B)$ explicitly, as well an upper bound on $\mathcal C(B)$. Moreover, for $d \geq 3$ in the limit of large regions, both $\mathcal H(B)$ and the bound on $\mathcal C(B)$ approach $\mathcal D_\text{eff}(B)$, the asymmetric domain of dependence defined by using $c_\text{eff}$ in place of $c$ for signal propagation on the brane. The EFT domain of dependence $\mathcal E_\alpha(B)$ is contained in the other regions for $\alpha \leq 1$. This establishes that all three regions are reasonable condidates for replacing the domain of dependence in KR models. The fact that $\mathcal E_\alpha(B)$ is contained in the other regions for $\alpha \leq 1$ means that, at least in the cases studied, the nice slice criterion might be sufficient to ensure that neither faster-than-light signalling is possible nor unitarity breaks down. 

In the following we will discuss the regions $\mathcal E_\alpha(B)$, $\mathcal{H}(B)$ and $\mathcal C(B)$ for the case where $B$ is a strip-shaped region. Recall that the metric in slicing coordinates is given by \cref{eq:SlicingMetric} with the brane located at $\mu=\mu_*$. The strip regions that are of interest to us have a finite extent in the $y$ direction ($x^1$ direction in the usual Poincar\'e coordinates, \cref{eq:poincare}) and will be infinite in all other transverse spatial directions. 

\subsection{Bounding the Unitary Domain of Dependence on the Brane}
\label{sec:Unitary_DoD_on_brane}
It is not clear how to evaluate the region $\mathcal C(B)$. However, it turns out that there is a relatively easy way to compute an upper bound on this region using subregion entropies.

In finite dimensional quantum systems we can associate entropies to subsystems. The natural notion in this case is the von Neumann entropy
\begin{align}
    S(B) = - \operatorname{tr}\left(\rho_B \log \rho_B\right).
\end{align}
As is evident from the definition, entropy is independent under the action of unitary operators on the reduced density matrix $\rho_B$ associated to $B$. In particular, this means that if we have two subsystems $B$ and $B'$ whose density matrices are related as in \cref{eq:unitary_time_evo} their entropy is the same. Moreover, entropy obeys a property known as \emph{subadditivity}. For any two, not necessarily disjoint, subsystems $X$ and $Y$ we have that
\begin{align}
    \label{eq:SA}
    S(X) + S(Y) \geq S(X \cup Y), &&\text{(subadditivity)}.
\end{align}

\begin{figure}
    \centering
    \begin{tikzpicture}
    \begin{scope}[scale=0.8]
        \draw (-3,0) node [below left] {$p_1$} -- node[below] {$B$} (3,0) node [below right] {$p_2$} -- node [above right] {$B_2$} (0.5,1.3) node [above] {$p_M$}  --  node [above left] {$B_1$} cycle;
    \end{scope}
    \end{tikzpicture}
    \caption{Configurations of subregions for which subadditivity is tested.}
    \label{fig:SSA_theo}
\end{figure}
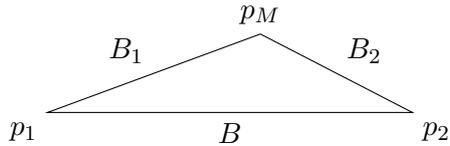

Our strategy to find an upper bound on $\mathcal C(B)$ is based on assessing the consistency of assigning entropies to subregions. To this end we study subadditivity on strip subregions of the brane. Our approach is summarized in \cref{fig:SSA_theo}. Consider a partial Cauchy slice $B$ at some time $t_0$. If we are inside the unitary domain of dependence \cref{eq:unitary_time_evo} ensures that there is a unitary operator $U$ with which we can evolve the state on $B$ to some other slice $B_1 \cup B_2$. We can then bound the region $\mathcal C(B)$ by moving the intersection $p_M = \partial B_1 \cap \partial B_2$ around and testing if \cref{eq:SA} holds for $X = B_1$ and $Y = B_2$. The region where subadditivity holds then places an upper bound on $\mathcal C(B)$, since failure of subadditivity indicates that our assumption that \cref{eq:unitary_time_evo} holds fails and we are outside $\mathcal C(B)$.

In quantum field theory, von Neumann entropy of subregions is ill-defined due to UV divergences. This seemingly renders \cref{eq:SA} useless. However, on the brane gravity is present and the natural notion of entropy is not given by the divergent von Neumann, but the finite generalized entropy. 
Generalized entropy is conjectured to be well-defined, i.e., finite and scheme-independent \cite{Susskind:1994sm} as divergences from $S_\text{matter}$ cancel against loop corrections to Newton's constant. Evidence that generalized entropy is indeed finite comes from holography for the cases where $B$ is an extremal surface \cite{Lewkowycz:2013nqa, Faulkner:2013ana}. While there is no equally compelling evidence in the case of $B$ being an arbitrary Cauchy splitting surface, the assumption that $S_\text{gen}$ is well-defined even in this situation underlies the quantum focusing conjecture \cite{Bousso:2015mna} and several corollaries, of which some
have been proven \cite{Bousso:2015wca, Balakrishnan:2017bjg}; see also \cite{Jensen:2023yxy,AliAhmad:2023etg} for recent discussions of generalized entropy of general subregions in the context of von Neumann algebras. For the purposes of this note we will simply assume that $S_\text{gen}$ is meaningful for arbitrary $B$ at leading order in $1/N$.

In practice this means that \cref{eq:SA} is in fact well defined if we consider the generalized entropy of subregions. Generalized entropy schematically takes the form
\begin{align}
\label{eq:sgen}
S_\text{gen}(B) = S_\text{geom}(B) + S_\text{matter}(B).
\end{align}
In semi-classical gravity, the first term $S_\text{geom}(B)$ is a geometric functional localized at the boundary of $B$ which depends on the gravity theory at hand.\footnote{Note that under certain circumstances the functional needs to be smeared over a region of the size of the cutoff, see e.g.\ \cite{Leichenauer:2017bmc}. In our case, however, the true geometric cutoff if given by the bulk string length, which is much smaller than the physical cutoff on the brane and thus we have no reason to do this. As we will see below, consistency does not depend on smearing.} As is well known, for Einstein gravity we have $S_\text{geom}(B) = \frac{\text{Area}(\partial B)}{4 G_N}$. The second term, $S_\text{matter}(B)$, computes the entropy of quantum fields in the region $B$.

In order to compute \cref{eq:sgen} we will use double holography and assume that $S_\text{gen}(B)$ on the brane is the area of the smallest bulk HRT surface \cite{Hubeny:2007xt,Ryu:2006bv} 
which is anchored on the brane at $\partial B$ divided by $4 G_N$.\footnote{In the context of $T\bar T$ deformed holographic theories it was similarly suggested that the HRT formula computes entropies in the deformed theory \cite{Donnelly:2018bef}.} Other possible holographic candidates like restricted minimax \cite{Grado-White:2020wlb} or causal holographic information \cite{Hubeny:2012wa,Freivogel:2013zta} are ruled out, because their values do not only depend on the entangling surface $\partial B$ as required by \cref{eq:unitary_time_evo}.
In addition to a lack of obvious alternative candidates, evidence for this prescription comes from the fact that extremal surfaces which end on the brane reproduce the leading order terms of the Wald-Dong entropy \cite{Wald:1993nt, Dong:2013qoa}. As mentioned in \cref{sec:KR_models} the gravity theory on the brane obtains higher curvature corrections, such that the geometric term in \cref{eq:sgen} takes the form
\begin{align}
    \label{eq:wald_dong}
     S_\text{geom}(B) = \frac 1 {4 G_N^{brane}} \left[ \int_{\partial B}\sqrt{h} + \frac {L^2} {2(d-4)(d-2)} \int_{\partial B} \sqrt{h} \left( 2 \mathcal R^a_a-\frac{d}{d-1} \mathcal R  - \mathcal K_a \mathcal K^a\right)\right] + \dots.
\end{align}
Computing the HRT surface and expanding it near the brane it can be shown that the contribution from an HRT surface precisely agrees with \cref{eq:wald_dong} \cite{Chen:2020uac}. This insight is crucial in doubly-holographic proofs of the island formula which prescribes how to compute the entropy of an evaporating black hole. This formula can be derived in two dimensions using Euclidean wormhole geometries, but can also be derived within double holography, provided that HRT surfaces which connect the asymptotic boundary to the brane compute generalized entropy in the brane picture.

Let us now turn to the explicit computation. Although slicing coordinates make the location of the brane very transparent, it is convenient to use the usual Poincar\'e coordinates on AdS, \cref{eq:poincare}, for the computation of entropy via the HRT prescription. The reason is that the equations that govern the HRT surfaces will be local and hence will have no knowledge about the presence of the brane. Thus, the equations for the HRT surfaces in this ETW brane setting will be identical to the ones in vacuum AdS. The only difference will be in the boundary conditions; the endpoints of the HRT surfaces lie on the brane in the ETW brane setting instead of the asymptotic AdS boundary.

Let us begin with an arbitrary $(d-1)$-dimensional (candidate HRT) surface in the bulk anchored to a strip region on the $d$-dimensional brane. We parametrize the surface with parameters $\zeta^0=z,\, \zeta^i=x^{i+1}~\text{for} ~ i=1,...,d-2$. The strip regions under consideration are infinitely extended in the $x^{i+1}$ directions (for $i=1,...,d-2$) implying that the $t$ and $x^1$ directions on the candidate HRT surface are independent of $\zeta^i$. In order to find the HRT surface to compute entropies, we need to extremize the area functional
\begin{equation}
    A=\int d^{d-1}\zeta\, \sqrt{\mathfrak{h}} ,
\end{equation}
where $\mathfrak{h}$ is the determinant of the induced metric on the candidate surface. The components of the induced metric are given by
\begin{equation}
\begin{split}
    \mathfrak{h}_{00}=L^2\left(\frac{1-t'(z)^2+{x^1}'(z)^2}{z^2}\right), \qquad \mathfrak{h}_{0i}=0, \qquad \mathfrak{h}_{ij}=\frac{L^2}{z^2}\delta_{ij},
\end{split}
\end{equation}
where the $'$ represents a derivative with respect to $\zeta^0=z$. With this induced metric the  problem reduces to extremizing 
\begin{equation}
\label{eqn:func_to_extremize}
    \frac{A}{V_{d-2}}=L^{d-1}\int^{z_f}_{z_i}  \frac{\sqrt{1-t'(z)^2+{x^1}'(z)^2}}{z^{d-1}}\,dz\,,
\end{equation}
with $V_{d-2}=\int d^{d-2}\zeta$ being the volume of the strip in the transverse directions and the limits of the integral ($z_i,\,z_f$) being fixed by the boundary of the strip region under analysis. The Euler-Lagrange equations are
\begin{equation}
\label{eqn:Euler_Lag}
    \begin{split}
        & \frac{t'(z)}{\sqrt{1-t'(z)^2+{x^1}'(z)^2}}=C_t\,z^{d-1}\\,
        & \frac{{x^1}'(z)}{\sqrt{1-t'(z)^2+{x^1}'(z)^2}}=C_x\,z^{d-1},
    \end{split}
\end{equation}
where $C_t$ and $C_x$ are integration constants which are again determined by the boundary of the strip region under analysis. We observe from \cref{eqn:Euler_Lag} that $dt/dx^1=C_t/C_x$ is a constant. Hence we define the boost associated to a partial Cauchy slice on the brane as
\begin{equation}
    \label{eqn:boost}
    \beta=\frac{\Delta t}{\Delta x^1}=\frac{t_f-t_i}{x^1_f-x^1_i}=\frac{C_t}{C_x},
\end{equation}
where the subscripts $f$ and $i$ indicate coordinates corresponding to the final and initial ending locations of the strip region.  Substituting for $t_z$ and $x^1_z$ from \cref{eqn:Euler_Lag} in \cref{eqn:func_to_extremize} we can get an expression for the entropy per unit transverse volume associated with a strip region as
\begin{equation}
\label{eqn:Strip_entropy}
\begin{split}
    \frac{S}{V_{d-2}}&=\frac{L^{d-1}}{4G_N}\int^{z_f}_{z_i}  \frac{z^{1-d}}{\sqrt{1-\left(C_x^2-C_t^2\right)z^{2(d-1)}}}\,dz\\
    &=\frac{L^{d-1}}{4G_N}\int^{z_f}_{z_i}  \frac{z^{1-d}}{\sqrt{1 \mp \left( z / z_t\right)^{2(d-1)}}}\,dz,
\end{split}
\end{equation}
 where $z_t=\left(\pm\left(C_x^2-C_t^2\right)\right)^{-1/2(d-1)}$ and the upper sign is used when $\beta^2<1$ and the lower sign otherwise. 

Note that if we perform the HRT surface calculation starting from a point on the imagined asymptotic boundary ($\mu=\pi/2$), for the cases when $\beta^2<1$, the HRT surface has a turning point in the bulk at $z=z_t$ and hence intersects the brane and the imagined boundary at two locations, i.e., at two different values of the coordinate $x^1$ each. In contrast, for the case when $\beta^2\geq1$, the HRT surface no longer has a turning point and intersects the imagined boundary at one location only. However, such an HRT surface may still intersect the brane at two locations and we also need to consider this case. We can get an analytical expression for the HRT surface profile and the entropy when we are in $2+1$ bulk dimensions (see \cref{sec:2+1_entropy}). 
In general, the above expressions \cref{eqn:Euler_Lag,eqn:Strip_entropy} can be integrated numerically to compute the HRT surface profile and the entropy associated to strip regions in arbitrary dimensions. With all the details spelled out, we can now use \cref{eqn:Strip_entropy} to compute entropies needed to analyse subadditivity, \cref{eq:SA}, on the brane.
\begin{figure}
    \centering
    \begin{tikzpicture}[scale=0.95]
\begin{axis}[
    title={$\mu_*=1.0$},
    xlabel={$y$},
    ylabel={$t$},
    xmin=1800, xmax=10200,
    ymin=0, ymax=4200,
    xtick={2000,4000,6000,8000,10000},
    ytick={0,1000,2000,3000,4000},
    scaled y ticks=base 10:-3,
    legend pos=north west,
    grid style=dashed,
        legend style={nodes={scale=0.7, transform shape}}
]
\addplot[
    color=black!90!white,dash dot, thick
    ] file {Data_Files/DofB_y1_2000_y2_10000.dat};
\addplot[
    color=black!50!white,dash dot,thick
    ] file {Data_Files/DeffB_mu_1p0_y1_2000_y2_10000.dat};
\addplot[color=teal, thick] file {Data_Files/Full_New_SubAd_d_3_mu_1p0_y1_2000_y2_10000.dat};

    \addlegendentry{$\partial^+\mathcal{D}(B)$};
    \addlegendentry{$\partial^+\mathcal{D}_\text{eff}(B)$};
    \addlegendentry{$\partial^+\mathcal{\Tilde{C}}(B)$};
    
\end{axis}
\end{tikzpicture}
\begin{tikzpicture}[scale=0.95]
\begin{axis}[
    title={$\mu_*=1.3$},
    xlabel={$y$},
    ylabel={$t$},
    xmin=1800, xmax=10200,
    ymin=0, ymax=4200,
    xtick={2000,4000,6000,8000,10000},
    ytick={0,1000,2000,3000,4000},
    scaled y ticks=base 10:-3,
    legend pos=north west,
    grid style=dashed,
    legend style={nodes={scale=0.7, transform shape}}
]
\addplot[
    color=black!90!white,dash dot, thick
    ] file {Data_Files/DofB_y1_2000_y2_10000.dat};
\addplot[
    color=black!50!white,dash dot, thick
    ] file {Data_Files/DeffB_mu_1p3_y1_2000_y2_10000.dat};
\addplot[color=teal, thick] file {Data_Files/Mod_Full_New_SubAd_d_3_mu_1p3_y1_2000_y2_10000.dat};
    \addlegendentry{$\partial^+\mathcal{D}(B)$};
    \addlegendentry{$\partial^+\mathcal{D}_\text{eff}(B)$};
    \addlegendentry{$\partial^+\mathcal{\Tilde{C}}(B)$};
    
\end{axis}
\end{tikzpicture}
    \caption{Comparing bounds on $\mathcal{C}(B)$ with $\mathcal{D}_\text{eff}(B)$ and $\mathcal{D}(B)$ for strip regions on $d=3$ dimensional branes ($y_1=2000$, $y_2=10000$). The AdS length scale $L$ and the bulk/boundary speed of light $c$ have been set to $1$. In this figure, as well as in all the following ones, we only plot $t >0$ since the plots are symmetric under time-reversal.}
    \label{fig:d2_UDoD_Compare}
\end{figure}
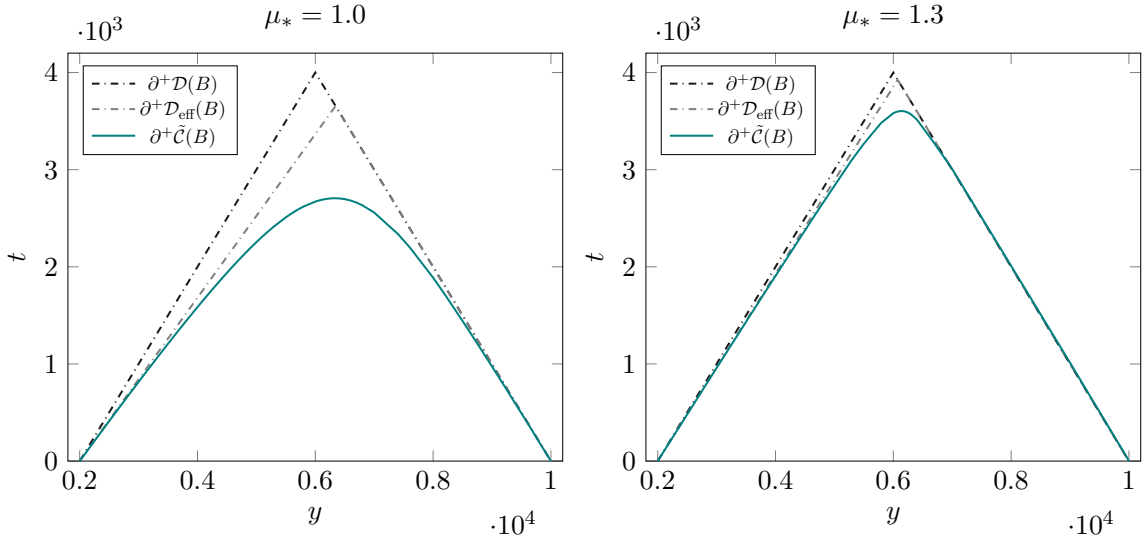

 Let us begin with a constant (Poincar\'e) time  partial Cauchy slice $B$ on the brane. To get an upper bound on the unitary domain of dependence $\mathcal{C}(B)$, we check subadditivity, \cref{eq:SA}, for strip regions on the brane in the configuration as in \cref{fig:SSA_theo}. Recall the slicing coordinates, \cref{eq:SlicingMetric}. We vary $y_M$ ($y$-coordinate of $p_M$) between $y_1$ ($y$-coordinate of $p_1$) and $y_2$ ($y$-coordinate of $p_2$) and find the maximum $t_M$ ($t$-coordinate of $p_M$), for each $y_M$, such that the subadditivity condition is satisfied. The curve traced by $t_M^\text{max}$ for different values of $y$ defines the boundary of a subregion on the brane where subadditivity holds. We will refer to this region as $\mathcal{\Tilde{C}}(B)$. 
 
 We illustrate the behavior of the future boundary of $\mathcal{\Tilde{C}}(B)$, denoted $\partial^+\mathcal{\Tilde{C}}(B)$, for a strip region on a $d=3$ dimensional brane in \cref{fig:d2_UDoD_Compare} (for plots corresponding to other values of $d$, see \cref{sec:Diff_DoD_comparison}). 
 
Subadditivity is not violated as long as $p_M$ is chosen from the region below the solid curve in \cref{fig:d2_UDoD_Compare}, whereas it is violated if $p_M$ is chosen from the region above the solid curve. 
As discussed earlier, this provides an upper bound on $\mathcal{C}(B)$. The future boundaries of $\mathcal{D}(B)$ and $\mathcal{D}_\text{eff}(B)$, denoted by $\partial^+\mathcal{D}(B)$ and $\partial^+ \mathcal{D}_\text{eff}(B)$ respectively, are shown as the dot-dashed lines in \cref{fig:d2_UDoD_Compare} for comparison. We observe that the upper bound on $\mathcal{C}(B)$ is a proper subset of $\mathcal{D}_\text{eff}(B)$ which in turn is contained in $\mathcal{D}(B)$. The past boundary $\partial^- \tilde C(B)$ can be obtained by realizing that our setup is time-reflection symmetric under $t \to -t$. We therefore only show $t>0$ in all plots. 
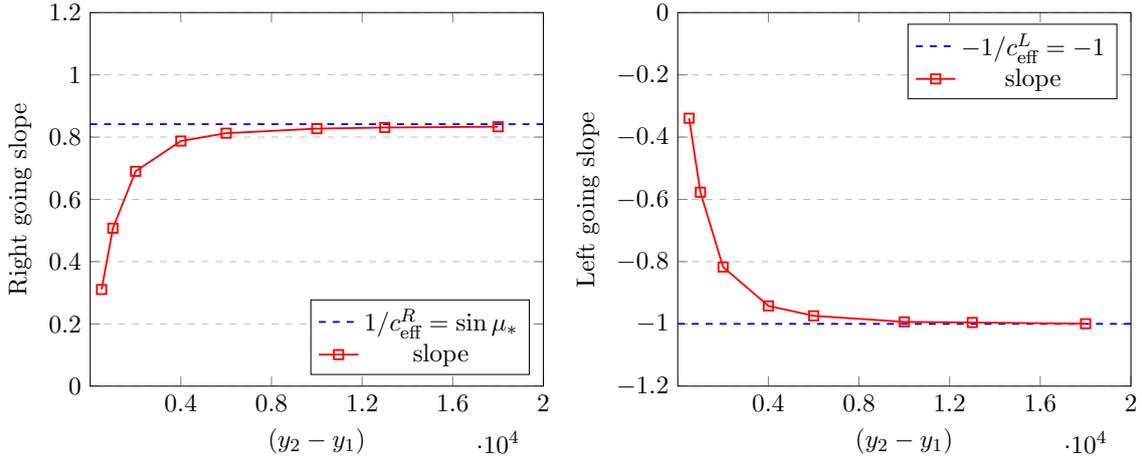
\begin{figure}
    \centering
\begin{tikzpicture}[scale=0.87]
\begin{axis}[
    xlabel={($y_2-y_1$)},
    ylabel={Right going slope},
    xmin=0, xmax=20000,
    ymin=0, ymax=1.2,
    xtick={4000,8000,12000,16000,20000},
    ytick={0,0.20,0.40,0.60,0.80,1.00,1.20},
    legend pos=south east,
    ymajorgrids=true,
    grid style=dashed,
]
\addplot[
    color=blue,dashed,thick
    ] file {Data_Files/Constant_sin_1p0.dat};
\addplot[
    color=blue,
    mark=square,
    ] [thick, red] file {Data_Files/Initial_slope_with_strip_region_size_mu_1p0_y1_2000_d_3.dat};
    \addlegendentry{$1/c^R_\text{eff}=\sin\mu_*$};
    \addlegendentry{slope};
\end{axis}
\end{tikzpicture}
\begin{tikzpicture}[scale=0.87]
\begin{axis}[
    xlabel={($y_2-y_1$)},
    ylabel={Left going slope},
    xmin=0, xmax=20000,
    ymin=-1.2, ymax=0,
    xtick={4000,8000,12000,16000,20000},
    ytick={0,-0.20,-0.40,-0.60,-0.80,-1.00,-1.20},
    legend pos=north east,
    ymajorgrids=true,
    grid style=dashed,
]
\addplot[
    color=blue,dashed,thick
    ] file {Data_Files/Constant_minus_one.dat};
\addplot[
    color=red,
    mark=square, thick
    ] file {Data_Files/End_slope_with_strip_region_size_mu_1p0_y1_2000_d_3.dat};
    \addlegendentry{$-1/c_\text{eff}^{L} = -1$};
    \addlegendentry{slope};
\end{axis}
\end{tikzpicture}
    \caption{Left: The slope of the SA bound at the end location closer to the defect approaches $1/c_\text{eff}^{R}=\sin\mu_* $ in the limit of large subregions. Right: The slope of the SA bound at the end location away from the defect approaches $-1/c_\text{eff}^{L}=-1$ in the limit of large subregions. These plots correspond to the case of a $d=3$ dimensional brane located at $\mu_*=1$. The AdS length scale $L$ and the bulk/boundary speed of light $c$ have been set to $1$.} 
    \label{fig:slope_comparison}
\end{figure}

An interesting observation about $\mathcal{\Tilde{C}}(B)$ is that the inverse slope of the solid curve in \cref{fig:d2_UDoD_Compare} at the end point closer to the defect is observed to be very close to the effective speed of light for propagation away from the defect ($c^R_\text{eff}=c/\sin\mu_*$) on the brane while the inverse slope at the end point away from the defect is observed to be very close to the effective speed of light for propagation towards the defect ($c^L_\text{eff}=c$). In fact, the inverse slopes converge to the respective speeds in the limit of large subregions for $d>2$.
This observation is illustrated in \cref{fig:slope_comparison} where we have plotted the slopes with increasing subregion size.\footnote{Note that the right going slope is the slope close to the (left) endpoint closer to the defect while the left going slope is the slope close to the (right) endpoint away from the defect.} This hints to the fact that $\mathcal{\Tilde{C}}(B)$ is always contained in $\mathcal{D}_\text{eff}(B)$ and approaches $\mathcal{D}_\text{eff}(B)$ in the limit of larger and larger subregions. 
Thus, $\mathcal{\Tilde{C}}(B)$ and hence $\mathcal{C}(B)$ always excludes the region affected by superluminal signalling on the brane reinforcing its candidacy for a viable replacement for the domain of dependence.

As we will see in \cref{fig:d2_EFTDoD_Compare}, the region $\tilde C(B)$ is symmetric for $d=2$, i.e., the left-moving effective speed of light is faster than $c$. This is not a problem, since we only require the alternative domains of dependence to be bounded by $\mathcal D_\text{eff}(B)$.
%
%
%

\subsection{The Holographic Domain of Dependence on the Brane}
Let us now turn to computing $\mathcal{H}(B)$ for constant time, strip-shaped regions $B$. Our main strategy will essentially be to compute the intersection of the brane with the light congruences that are normal to the HRT surface. There are two subtleties however. First, in higher dimensions we need to be careful to truncate light rays which end in a caustic. Second, we are of course only interested in light congruences which originate at the HRT surface in the physical part of the spacetime, i.e., in ``front'' of the brane. If the angle between the brane and the HRT surface is bigger than $\frac \pi 2$, which can happen for the end of the strip further away from the defect, light rays which are normal to the HRT surface do not fully enclose the entanglement wedge and do not necessarily intersect the brane. As already discussed in \cref{sec:a_puzzle}, in this case we need to add a light cone at the point where the HRT surface ends on the brane to complete the lightsheet that bounds the entanglement wedge. Thus, if the angle between the HRT surface and the brane is bigger than $\frac \pi 2$, the intersection of the entanglement wedge and the brane will be completed by a $c=1$ light sheet on the brane propagating towards the defect, in accordance with \cref{sec:speed_of_light}.

For the case of $d=2$ dimensional brane, the calculation of $\mathcal{H}(B)$ is extremely straightforward (due to the absence of caustics and the fact that the intersection angle of the HRT surface with the brane is always less than $\frac \pi 2$) and reduces to simply computing the intersection of the brane with a cone, see \cref{sec:Ent_intersection_brane}. In general dimensions, the easiest way to perform the required computation is again to go to Poincar\'e coordinates, \cref{eq:poincare}, which cover the bulk spacetime including an imagined part behind the brane up to $z=0$, where we have an imagined boundary. Given Poincar\'e coordinates $t, \vec x, z$ we will use $x^{i}$, $i \geq 2$ and $z$ to parametrize the HRT surface via $(t=0, x^1 = x^1(z), x^2, \dots, z)$. Recall that here we have taken the finite width of the strip is in the $x^1$ direction. To find the corresponding entanglement wedge, we are interested in light rays that propagate towards the (imagined) asymptotic boundary. The trajectory of a light ray with $c=1$ can be parametrized by Poincar\'e time $t$ as 
\begin{align}\label{eq:lightray}
    l^\mu(t) = \begin{pmatrix} t \\ (t_0 - t) \sin \theta \sin \Omega_i
 + x^1_{0} \\(t_0 - t) \sin \theta \cos \Omega_i + x^2_{0} \\ \dots \\ (t_0 - t) \cos \theta \end{pmatrix},
\end{align}
where $t_0$, $x_0^i$ determine the location where the light ray intersects the asymptotic boundary and $\theta$, $\Omega_i$ determine the direction.

Requiring that this trajectory lies in the normal plane of the HRT surface at $t=0, x^1 = x^1(z), x^2, \dots z$ yields
\begin{align}
    \cos \Omega_i = 0, && \cos \theta = \frac{\pm {x^1}'(z)}{\sqrt{1 + {x^1}'(z)^2}}.
\end{align}
The upper sign must be chosen if ${x^1}'(z)$ is positive, otherwise we have to take the lower sign.

It now remains to fix the remaining constants $t_0, x^i_{0}, \dots$ in \cref{eq:lightray} by requiring that \cref{eq:lightray} passes through the point of the HRT surface that is parameterized by $x^2, x^3, \dots, z$. The result is
\begin{align}
\begin{split}
    t_0 &= \pm \frac{z \sqrt{1 + {x^1}'(z)^2}}{{x^1}'(z)},\\
    x^1_{0} &= x^1(z) + \frac{z}{{x^1}'(z)},  \\ 
    x^i_{0} &= x^i, \qquad \qquad (\text{for }i \geq 2).
\end{split}
\end{align}
Substituting this back into \cref{eq:lightray}, we obtain an expression for the light ray that starts at the HRT surface at point $(t=0, x^1 = x^1(z), x^2, \dots z)$, is normal to the HRT surface and propagates towards the boundary,
\begin{align}
    l^\mu(t|x_2, \dots ,z) = \begin{pmatrix} t \\ x(z) \pm \frac{t}{\sqrt{1+{x^1}'(z)^2}}       \\ x_2 \\ \vdots \\ z \mp t\frac{{x^1}'(z)}{\sqrt{1+{x^1}'(z)^2}} \end{pmatrix},
\end{align}
parametrized by Poincar\'e time $t$.

It is now easy to compute the intersection of the lightray with the brane. The light ray intersects the brane when $l^{1} = l^{d} \tan \mu_*$. This can be solved for $t$ and one finds
\begin{align}
     t = \pm \sigma(z) \sqrt{1 + {x^1}'(z)^2 } ,
\end{align}
where we have defined
\begin{align}
    \sigma(z) = \frac{\sin \mu_* \; z - \cos \mu_* \; x^1(z)}{\sin \mu_* \; {x^1}'(z) + \cos \mu_*}.
\end{align}
Substituting this back into the parametrized form of the light ray determines the intersection locus
\begin{align}
\label{eq:intersection}
l_\cap^\mu(x^2, x^3, \dots,z) = \begin{pmatrix} 
\pm \sigma(z) \sqrt{1 + {x^1}'(z)^2 }  \\ 
x^1(z) + \sigma(z) \\ 
x^2 \\
\vdots\\
z - \sigma(z) {x^1}'(z) \end{pmatrix}.
\end{align}
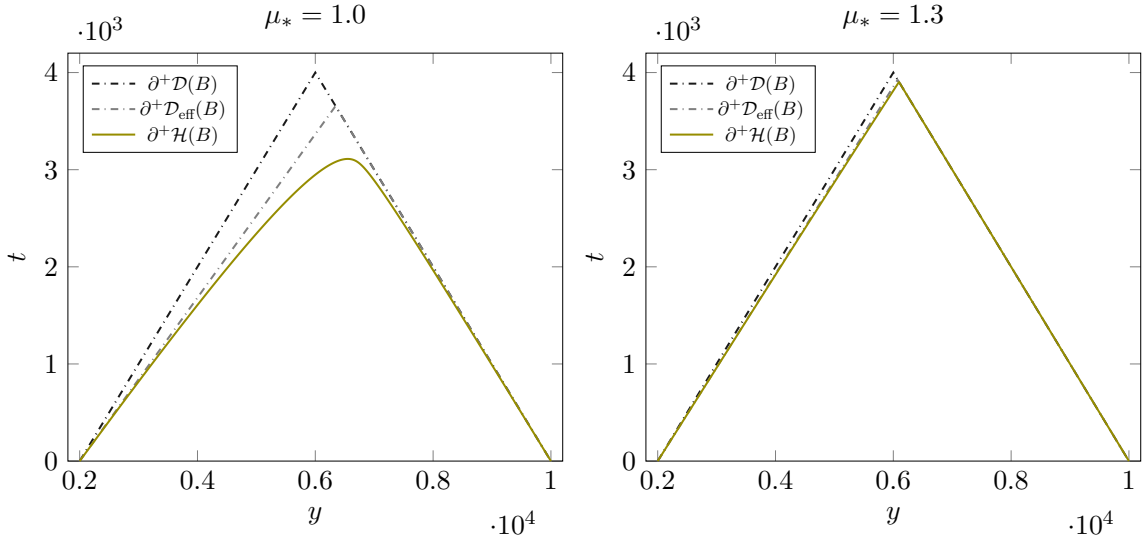
\begin{figure}
    \centering
    \begin{tikzpicture}[scale=0.95]
\begin{axis}[
    title={$\mu_*=1.0$},
    xlabel={$y$},
    ylabel={$t$},
    xmin=1800, xmax=10200,
    ymin=0, ymax=4200,
    xtick={2000,4000,6000,8000,10000},
    ytick={0,1000,2000,3000,4000},
    scaled y ticks=base 10:-3,
    legend pos=north west,
    grid style=dashed,
        legend style={nodes={scale=0.7, transform shape}}
]
\addplot[
    color=black!90!white,dash dot, thick
    ] file {Data_Files/DofB_y1_2000_y2_10000.dat};
\addplot[
    color=black!50!white,dash dot,thick
    ] file {Data_Files/DeffB_mu_1p0_y1_2000_y2_10000.dat};
\addplot[color=olive, thick] file {Data_Files/Ent_Wd_Int_brane_mu_1p0_y1_2000_y2_10000_d_3.dat};

    \addlegendentry{$\partial^+\mathcal{D}(B)$};
    \addlegendentry{$\partial^+\mathcal{D}_\text{eff}(B)$};
    \addlegendentry{$\partial^+\mathcal{H}(B)$};
    
\end{axis}
\end{tikzpicture}
\begin{tikzpicture}[scale=0.95]
\begin{axis}[
    title={$\mu_*=1.3$},
    xlabel={$y$},
    ylabel={$t$},
    xmin=1800, xmax=10200,
    ymin=0, ymax=4200,
    xtick={2000,4000,6000,8000,10000},
    ytick={0,1000,2000,3000,4000},
    scaled y ticks=base 10:-3,
    legend pos=north west,
    grid style=dashed,
    legend style={nodes={scale=0.7, transform shape}}
]
\addplot[
    color=black!90!white,dash dot, thick
    ] file {Data_Files/DofB_y1_2000_y2_10000.dat};
\addplot[
    color=black!50!white, dash dot, thick
    ] file {Data_Files/DeffB_mu_1p3_y1_2000_y2_10000.dat};
\addplot[color=olive, thick] file {Data_Files/Mod_Ent_Wd_Int_brane_mu_1p3_y1_2000_y2_10000_d_3.dat};
    \addlegendentry{$\partial^+\mathcal{D}(B)$};
    \addlegendentry{$\partial^+\mathcal{D}_\text{eff}(B)$};
    \addlegendentry{$\partial^+\mathcal{H}(B)$};

\end{axis}
\end{tikzpicture}
    \caption{Comparing $\mathcal{H}(B)$, $\mathcal{D}_\text{eff}(B)$ and $\mathcal{D}(B)$ for strip regions on $d=3$ dimensional branes ($y_1=2000$, $y_2=10000$). The AdS length scale $L$ and the bulk/boundary speed of light $c$ have been set to $1$. }
    \label{fig:d2_HDoD_Compare}
\end{figure}
Lastly we relate this to coordinates on the brane using \cref{eqn:Slicing_to_poincare}. Since the HRT surface sits at constant $x^i$, $i \geq 2$, the resulting intersection of the entanglement wedge with the brane is shift symmetric along the brane directions $\xi^i$. The non-trivial profile is captured by $l^0_\cap$ and $l^{d+1}_\cap$. It is most convenient to express it parametrized by $z$. In brane coordinates it reads
\begin{align}
    \label{eq:parametric}
    \begin{pmatrix} t_\cap(z) \\ y_\cap(z) \end{pmatrix} = \begin{pmatrix} \pm \sigma(z) \sqrt{1 + {x^1}'(z)^2 }  \\ \frac{z + x^1(z) {x^1}'(z)}{\sin \mu_* \; {x^1}'(z) + \cos \mu_*}\end{pmatrix}.
\end{align}
This expression is already sufficient to show that the holographic domain of dependence is bounded by a signal that propagates at $c^R_\text{eff}$ away from the brane. To see this, we can compute the slope of the curve defined by \cref{eq:parametric} for ${x^1}'(z)\geq 0$ ,
\begin{align}
    \frac{t'_\cap(z)}{y'_\cap(z)} = \frac {\sin \mu_* - \cos \mu_* \;{x^1}'(z)} {\sqrt{1 + {x^1}'(z)^2}} \leq \sin \mu_* = \frac 1 {c_\text{eff}^R}.
\end{align}
The comparison of the holographic domain of dependence $\mathcal{H}(B)$ with $\mathcal{D}_\text{eff}(B)$ and $\mathcal{D}(B)$ for a strip region on a $d=3$ dimensional brane is illustrated in \cref{fig:d2_HDoD_Compare}. There, we can also visually confirm that the future boundary of $\mathcal{H}(B)$, denoted $\partial^+\mathcal{H}(B)$ closer to the defect is bounded by $c_\text{eff}^R$ and thus $\mathcal H(B)$ does not include the region affected by superluminal signalling on the brane.


\subsection{The EFT Domain of Dependence on the Brane}
\label{sec:EFT_DoD}
Finally, let us calculate $\mathcal{E}_{\alpha}(B)$ for strip regions on the brane.
In particular, for a constant time Cauchy slice on the brane like the strip region $B$ in \cref{fig:SSA_theo}, the future boundary of $\mathcal{E}_{\alpha}(B)$ will again be a partial Cauchy slice $\Tilde{B_{\alpha}}$ with $\partial \Tilde{B_{\alpha}}=\partial B$ such that the extrinsic curvature scalar $\mathfrak{K}(\Tilde{B_{\alpha}})=\alpha/\Lcut$, where $\Lcut=L/\sin\mu_*$ is the cutoff scale for the theory on the brane as discussed in \cref{sec:scale_of_cutoff} (also see \cref{fig:EFT_DoD_fig}).

We will restrict ourselves to computing $\mathfrak{K}$ for a strip region on the brane, since as noted before the scale of extrinsic and intrinsic curvature are the same in the case at hand. 
Recall the slicing coordinates, \cref{eq:SlicingMetric}. We will parametrize the strip regions on the brane with parameters $\zeta^0=y,\, \zeta^i=\xi^{i+1}~\text{for} ~ i=1,...,d-2$. For a strip region with arbitrary $t=f(y)$ profile (infinitely extended in the transverse directions), the components the extrinsic curvature $\mathfrak{K}_{ab}$ for the strip region are given by
\begin{equation}
    \begin{split}
        & \mathfrak{K}_{00}=-L\,\frac{yf''(y)-f'(y)+(f'(y))^3}{y^2 \cos(\mu_*)\sqrt{1-(f'(y))^2}},\\
        & \mathfrak{K}_{0i}=0,\\
        & \mathfrak{K}_{ij}=L\,\frac{f'(y)}{y^2 \cos(\mu_*)\sqrt{1-(f'(y))^2}} \delta_{ij},
    \end{split}
\end{equation}
where $\mu=\mu_*$ describes the location of the brane in slicing coordinates. The extrinsic curvature scalar is given by
\begin{equation}
\label{eqn:strip_ext_curv_scalar}
    \mathfrak{K}=-\cos(\mu_*)\,\frac{y f''(y)+(d-1)(f'(y)^3-f'(y))}{\left(1-f'(y)^2\right)^{3/2}}.
\end{equation}
Solving for $\mathfrak{K}=\alpha\sin\mu_*/L$ with the boundary conditions same as that of the slice $B$ will give us the partial Cauchy slices $\Tilde{B_{\alpha}}$ that form the future boundary of $\mathcal{E}_{\alpha}(B)$ (see \cref{fig:EFT_DoD_fig}). For $d=2$, an analytical solution for $\mathfrak{K}=\sin\mu_*/L$ is given by
\begin{equation}
\label{eqn:Ent_wedge_int_brane}
    f(y)=\frac{1}{2\sin\mu_*}\left(\sqrt{y_1^2+y_2^2+2y_1y_2\cos(2\mu_*)}-\sqrt{y^2\sin^2(2\mu_*)+\left(y_1+y_2-2y\sin^2\mu_*\right)^2}\right),
\end{equation}
where $y_1,\, y_2$ are the $y$ coordinates of the end points $p_1$ and $p_2$ of the partial Cauchy slice as in \cref{fig:EFT_DoD_fig}. For other generic values of $\mathfrak{K}$ for $d=2$ and for $d>2$, we solve for $f(y)$ in \cref{eqn:strip_ext_curv_scalar} numerically.

\begin{figure}
    \centering
    \begin{tikzpicture}[scale=0.95]
\begin{axis}[
    title={$\mu_*=1.0$},
    xlabel={$y$},
    ylabel={$t$},
    xmin=1800, xmax=10200,
    ymin=0, ymax=4200,
    xtick={2000,4000,6000,8000,10000},
    ytick={0,1000,2000,3000,4000},
    scaled y ticks=base 10:-3,
    legend pos=north west,
    grid style=dashed,
        legend style={nodes={scale=0.6, transform shape}}
]
\addplot[
    color=black!90!white,dash dot, thick
    ] file {Data_Files/DofB_y1_2000_y2_10000.dat};
\addplot[
    color=black!50!white,dash dot,thick
    ] file {Data_Files/DeffB_mu_1p0_y1_2000_y2_10000.dat};

\addplot[color=black!60!red, thick] file {Data_Files/Const_Ext_Curv_alpha_1_mu_1p0_y1_2000_y2_10000_d_3.dat};
\addplot[color=black!30!red, thick] file {Data_Files/Const_Ext_Curv_alpha_p5_mu_1p0_y1_2000_y2_10000_d_3.dat};
\addplot[color=red,thick] file {Data_Files/Const_Ext_Curv_alpha_p1_mu_1p0_y1_2000_y2_10000_d_3.dat};
    \addlegendentry{$\partial^+\mathcal{D}(B)$};
    \addlegendentry{$\partial^+\mathcal{D}_\text{eff}(B)$};
    \addlegendentry{$\partial^+\mathcal{E}_{1}(B)$};
    \addlegendentry{$\partial^+\mathcal{E}_{0.5}(B)$};
    \addlegendentry{$\partial^+\mathcal{E}_{0.1}(B)$};

\end{axis}
\end{tikzpicture}
\begin{tikzpicture}[scale=0.95]
\begin{axis}[
    title={$\mu_*=1.3$},
    xlabel={$y$},
    ylabel={$t$},
    xmin=1800, xmax=10200,
    ymin=0, ymax=4200,
    xtick={2000,4000,6000,8000,10000},
    ytick={0,1000,2000,3000,4000},
    scaled y ticks=base 10:-3,
    legend pos=north west,
    grid style=dashed,
    legend style={nodes={scale=0.6, transform shape}}
]
\addplot[
    color=black!90!white,dash dot, thick
    ] file {Data_Files/DofB_y1_2000_y2_10000.dat};
\addplot[
    color=black!50!white, dash dot, thick
    ] file {Data_Files/DeffB_mu_1p3_y1_2000_y2_10000.dat};
\addplot[color=black!60!red, thick] file {Data_Files/Const_Ext_Curv_alpha_1_mu_1p3_y1_2000_y2_10000_d_3.dat};
\addplot[color=black!30!red, thick] file {Data_Files/Const_Ext_Curv_alpha_p5_mu_1p3_y1_2000_y2_10000_d_3.dat};
\addplot[color=red,thick] file {Data_Files/Const_Ext_Curv_alpha_p1_mu_1p3_y1_2000_y2_10000_d_3.dat};
    \addlegendentry{$\partial^+\mathcal{D}(B)$};
    \addlegendentry{$\partial^+\mathcal{D}_\text{eff}(B)$};
    \addlegendentry{$\partial^+\mathcal{E}_{1}(B)$};
    \addlegendentry{$\partial^+\mathcal{E}_{0.5}(B)$};
    \addlegendentry{$\partial^+\mathcal{E}_{0.1}(B)$};
    
\end{axis}
\end{tikzpicture}
    \caption{Comparing $\mathcal{E}_{\alpha}(B)$, $\mathcal{D}_\text{eff}(B)$ and $\mathcal{D}(B)$ for strip regions on $d=3$ dimensional branes ($y_1=2000$, $y_2=10000$). The AdS length scale $L$ and the bulk/boundary speed of light $c$ have been set to $1$. }
    \label{fig:d2_EFTDoD_Compare}
\end{figure}
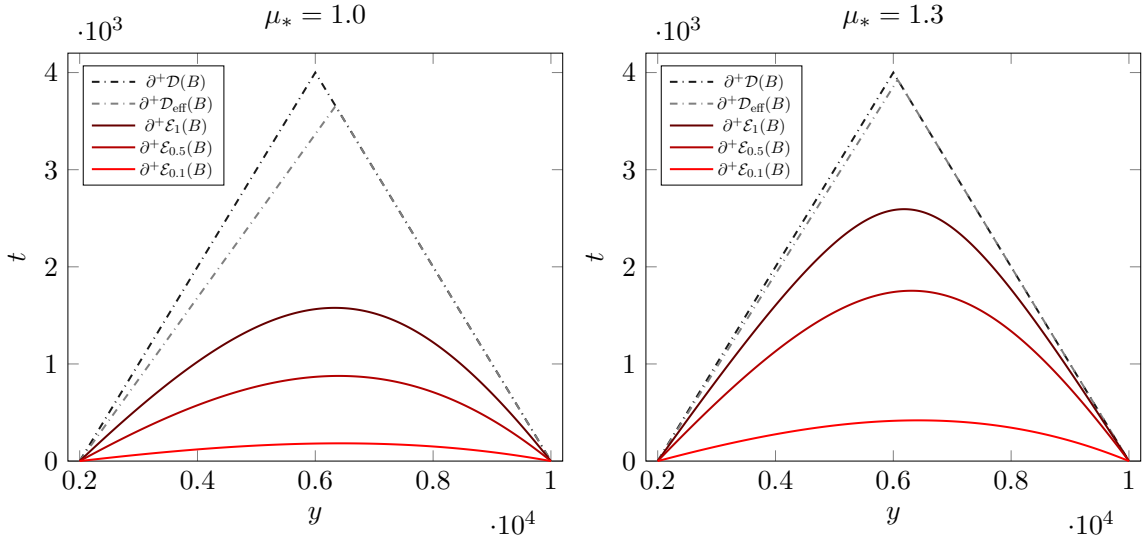
The $t-y$ cross-sections of a Cauchy slice for which the extrinsic curvature scalar equals to the cutoff scale on the brane ($\mathfrak{K}=\alpha \sin\mu_*/L$) on a $d=3$ dimensional brane is illustrated in \cref{fig:d2_EFTDoD_Compare}.
These are the future boundaries of the EFT domains of dependence $\mathcal{E}_{\alpha}(B)$ (denoted $\partial^+ \mathcal{E}_{\alpha}(B)$). Any partial Cauchy slice $B'$ (with $\partial B'=\partial B$) having a $t-y$ cross-section above the solid lines in \cref{fig:d2_EFTDoD_Compare} is not a nice slice.
Comparing with $\mathcal{D}_\text{eff}(B)$ and $\mathcal{D}(B)$, we observe that the EFT domain of dependence $\mathcal{E}_{\alpha}(B)$ is contained in $\mathcal{D}_\text{eff}(B)$ and hence excludes the regions affected by superluminal signalling on the brane. It is curious to note that the constant extrinsic curvature slices are slightly asymmetrical with respect to the coordinate center of the partial Cauchy slice in higher dimensions (\cref{fig:d2_EFTDoD_Compare,fig:DoD_Compare_3-4D}) whereas there is no such asymmetry for the case of $d=2$ dimensional branes (\cref{fig:d2_DoD_Compapre}). 
\subsection{Comparing  $\mathcal{H}$, $\mathcal{E}_{\alpha}$ and the Upper Bound on $\mathcal{C}$}
\label{sec:Diff_DoD_comparison}

\begin{figure}
    \centering
    \begin{tikzpicture}[scale=0.95]
\begin{axis}[
    title={$\mu_*=1.0$},
    xlabel={$y$},
    ylabel={$t$},
    xmin=1800, xmax=10200,
    ymin=0, ymax=4200,
    xtick={2000,4000,6000,8000,10000},
    ytick={0,1000,2000,3000,4000},
    scaled y ticks=base 10:-3,
    legend pos=north west,
    grid style=dashed,
        legend style={nodes={scale=0.54, transform shape}}
]
\addplot[
    color=black!90!white,dash dot, thick
    ] file {Data_Files/DofB_y1_2000_y2_10000.dat};
    \addplot[
    color=black!50!white,dash dot, thick
    ] file {Data_Files/DeffB_mu_1p0_y1_2000_y2_10000.dat};
\addplot[
    color=olive, dotted, line width=0.8mm
    ] file {Data_Files/Ent_Wd_Int_brane_mu_1p0_y1_2000_y2_10000_d_2.dat};
\addplot[
    color=teal,line width=0.4mm
    ] file {Data_Files/SA_bound_d_2_mu_1p0_y1_2000_y2_10000.dat};

\addplot[
    color=black!60!red,dashed,line width=0.6mm
    ] file {Data_Files/Const_Ext_Curv_alpha_1_mu_1p0_y1_2000_y2_10000_d_2.dat};
    \addplot[
    color=black!30!red,dashed,line width=0.4mm
    ] file {Data_Files/Const_Ext_Curv_alpha_p5_mu_1p0_y1_2000_y2_10000_d_2.dat};
    \addplot[
    color=red,dashed,line width=0.4mm
    ] file {Data_Files/Const_Ext_Curv_alpha_p1_mu_1p0_y1_2000_y2_10000_d_2.dat};
    \addlegendentry{$\partial^+\mathcal{D}(B)$};
    \addlegendentry{$\partial^+\mathcal{D}_\text{eff}(B)$};
    \addlegendentry{$\partial^+\mathcal{H}(B)$};
    \addlegendentry{$\partial^+\mathcal{\Tilde{C}}(B)$};
    \addlegendentry{$\partial^+\mathcal{E}_{1}(B)$};
    \addlegendentry{$\partial^+\mathcal{E}_{0.5}(B)$};
    \addlegendentry{$\partial^+\mathcal{E}_{0.1}(B)$};

\end{axis}
\end{tikzpicture}
\begin{tikzpicture}[scale=0.95]
\begin{axis}[
    title={$\mu_*=1.3$},
    xlabel={$y$},
    ylabel={$t$},
    xmin=1800, xmax=10200,
    ymin=0, ymax=4200,
    xtick={2000,4000,6000,8000,10000},
    ytick={0,1000,2000,3000,4000},
    scaled y ticks=base 10:-3,
    legend pos=north west,
    grid style=dashed,
    legend style={nodes={scale=0.54, transform shape}}
]
\addplot[
    color=black!90!white,dash dot, thick
    ] file {Data_Files/DofB_y1_2000_y2_10000.dat};
\addplot[
    color=black!50!white,dash dot, thick
    ] file {Data_Files/DeffB_mu_1p3_y1_2000_y2_10000.dat};
\addplot[
    color=olive, dotted, line width=0.8mm
    ] file {Data_Files/Ent_Wd_Int_brane_mu_1p3_y1_2000_y2_10000_d_2.dat};
\addplot[
    color=teal, line width=0.4mm
    ] file {Data_Files/SA_bound_d_2_mu_1p3_y1_2000_y2_10000.dat};
    \addplot[
    color=black!60!red,dashed,line width=0.6mm
    ] file {Data_Files/Const_Ext_Curv_alpha_1_mu_1p3_y1_2000_y2_10000_d_2.dat};
    \addplot[
    color=black!30!red,dashed,line width=0.4mm
    ] file {Data_Files/Const_Ext_Curv_alpha_p5_mu_1p3_y1_2000_y2_10000_d_2.dat};
    \addplot[
    color=red,dashed,line width=0.4mm
    ] file {Data_Files/Const_Ext_Curv_alpha_p1_mu_1p3_y1_2000_y2_10000_d_2.dat};
    \addlegendentry{$\partial^+\mathcal{D}(B)$};
    \addlegendentry{$\partial^+\mathcal{D}_\text{eff}(B)$};
    \addlegendentry{$\partial^+\mathcal{H}(B)$};
    \addlegendentry{$\partial^+\mathcal{\Tilde{C}}(B)$};
    \addlegendentry{$\partial^+\mathcal{E}_{1}(B)$};
    \addlegendentry{$\partial^+\mathcal{E}_{0.5}(B)$};
    \addlegendentry{$\partial^+\mathcal{E}_{0.1}(B)$};
    
\end{axis}
\end{tikzpicture}
    \caption{Comparing $\mathcal{H}(B)$ and $\mathcal{E}_{\alpha}(B)$ and bound on $\mathcal{C}(B)$ for strip regions on $d=2$ dimensional branes ($y_1=2000$, $y_2=10000$). The constant-$\mathfrak K$ curve for $\alpha=1$, the SA bound and the intersection of the EW with the brane all coincide exactly. The AdS length scale $L$ and the bulk/boundary speed of light $c$ have been set to $1$. }
    \label{fig:d2_DoD_Compapre}
\end{figure}
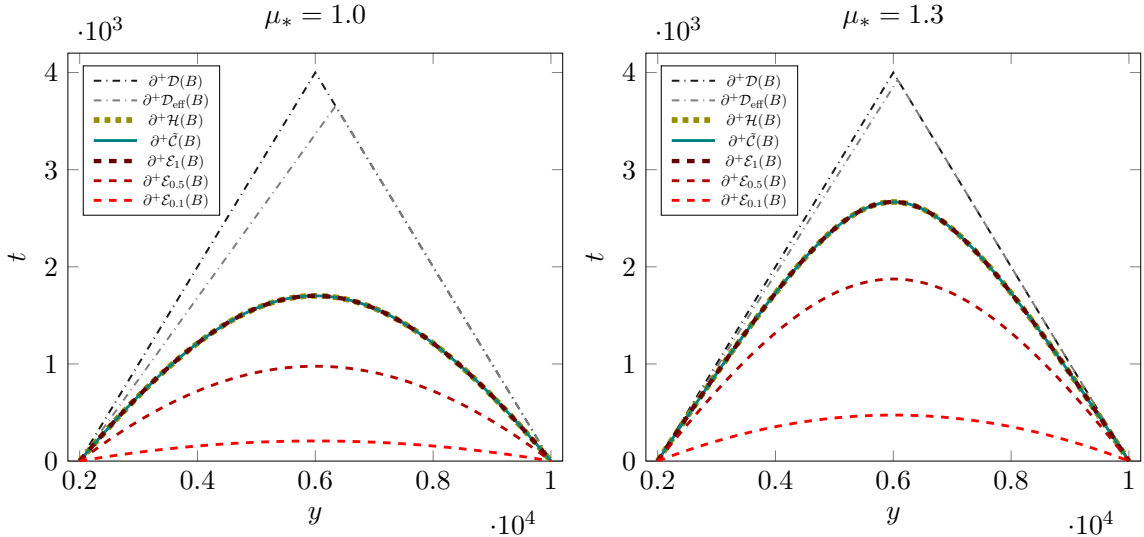
\begin{figure}
    \centering
    \begin{tikzpicture}[scale=0.95]
\begin{axis}[
    title={$d=3$},
    xlabel={$y$},
    ylabel={$t$},
    xmin=1800, xmax=10200,
    ymin=0, ymax=4200,
    xtick={2000,4000,6000,8000,10000},
    ytick={0,1000,2000,3000,4000},
    scaled y ticks=base 10:-3,
    legend pos=north west,
    grid style=dashed,
        legend style={nodes={scale=0.6, transform shape}}
]
\addplot[
    color=olive, dotted, line width=0.4mm
    ] file {Data_Files/Ent_Wd_Int_brane_mu_1p0_y1_2000_y2_10000_d_3.dat};
\addplot[
    color=teal,line width=0.4mm
    ] file {Data_Files/Full_New_SubAd_d_3_mu_1p0_y1_2000_y2_10000.dat};
    \addplot[
    color=black!60!red,dashed,line width=0.4mm
    ] file {Data_Files/Const_Ext_Curv_alpha_1_mu_1p0_y1_2000_y2_10000_d_3.dat};
    \addplot[
    color=black!30!red,dashed,line width=0.4mm
    ] file {Data_Files/Const_Ext_Curv_alpha_p5_mu_1p0_y1_2000_y2_10000_d_3.dat};
    \addplot[
    color=red,dashed,line width=0.4mm
    ] file {Data_Files/Const_Ext_Curv_alpha_p1_mu_1p0_y1_2000_y2_10000_d_3.dat};

    \addlegendentry{$\partial^+\mathcal{H}(B)$};
    \addlegendentry{$\partial^+\mathcal{\Tilde{C}}(B)$};
    \addlegendentry{$\partial^+\mathcal{E}_{1}(B)$};
    \addlegendentry{$\partial^+\mathcal{E}_{0.5}(B)$};
    \addlegendentry{$\partial^+\mathcal{E}_{0.1}(B)$};

\end{axis}
\end{tikzpicture}
\begin{tikzpicture}[scale=0.95]
\begin{axis}[
    title={$d=4$},
    xlabel={$y$},
    ylabel={$t$},
    xmin=1800, xmax=10200,
    ymin=0, ymax=4200,
    xtick={2000,4000,6000,8000,10000},
    ytick={0,1000,2000,3000,4000},
    scaled y ticks=base 10:-3,
    legend pos=north west,
    grid style=dashed,
    legend style={nodes={scale=0.6, transform shape}}
]

\addplot[
    color=olive, dotted, line width=0.4mm
    ] file {Data_Files/Mod_Ent_Wd_Int_brane_mu_1p0_y1_2000_y2_10000_d_4.dat};
\addplot[
    color=teal, line width=0.4mm
    ] file {Data_Files/Full_New_SubAd_d_4_mu_1p0_y1_2000_y2_10000.dat};
    \addplot[
    color=black!60!red,dashed,line width=0.4mm
    ] file {Data_Files/Const_Ext_Curv_alpha_1_mu_1p0_y1_2000_y2_10000_d_4.dat};
    \addplot[
    color=black!30!red,dashed,line width=0.4mm
    ] file {Data_Files/Const_Ext_Curv_alpha_p5_mu_1p0_y1_2000_y2_10000_d_4.dat};
    \addplot[
    color=red,dashed,line width=0.4mm
    ] file {Data_Files/Const_Ext_Curv_alpha_p1_mu_1p0_y1_2000_y2_10000_d_4.dat};

    \addlegendentry{$\partial^+\mathcal{H}(B)$};
    \addlegendentry{$\partial^+\mathcal{\Tilde{C}}(B)$};
    \addlegendentry{$\partial^+\mathcal{E}_{1}(B)$};
    \addlegendentry{$\partial^+\mathcal{E}_{0.5}(B)$};
    \addlegendentry{$\partial^+\mathcal{E}_{0.1}(B)$};
    
\end{axis}
\end{tikzpicture}
    \begin{tikzpicture}[scale=0.95]
\begin{axis}[
    xlabel={$y$},
    ylabel={$t$},
    xmin=1800, xmax=10200,
    ymin=0, ymax=4200,
    xtick={2000,4000,6000,8000,10000},
    ytick={0,1000,2000,3000,4000},
    scaled y ticks=base 10:-3,
    legend pos=north west,
    grid style=dashed,
        legend style={nodes={scale=0.6, transform shape}}
]
\addplot[
    color=olive, dotted, line width=0.4mm
    ] file {Data_Files/Mod_Ent_Wd_Int_brane_mu_1p3_y1_2000_y2_10000_d_3.dat};
\addplot[
    color=teal,line width=0.4mm
    ] file {Data_Files/Mod_Full_New_SubAd_d_3_mu_1p3_y1_2000_y2_10000.dat};
    \addplot[
    color=black!60!red,dashed,line width=0.4mm
    ] file {Data_Files/Const_Ext_Curv_alpha_1_mu_1p3_y1_2000_y2_10000_d_3.dat};
    \addplot[
    color=black!30!red,dashed,line width=0.4mm
    ] file {Data_Files/Const_Ext_Curv_alpha_p5_mu_1p3_y1_2000_y2_10000_d_3.dat};
    \addplot[
    color=red,dashed,line width=0.4mm
    ] file {Data_Files/Const_Ext_Curv_alpha_p1_mu_1p3_y1_2000_y2_10000_d_3.dat};
    \addlegendentry{$\partial^+\mathcal{H}(B)$};
    \addlegendentry{$\partial^+\mathcal{\Tilde{C}}(B)$};
    \addlegendentry{$\partial^+\mathcal{E}_{1}(B)$};
    \addlegendentry{$\partial^+\mathcal{E}_{0.5}(B)$};
    \addlegendentry{$\partial^+\mathcal{E}_{0.1}(B)$};
    
\end{axis}
\end{tikzpicture}
\begin{tikzpicture}[scale=0.95]
\begin{axis}[
    xlabel={$y$},
    ylabel={$t$},
    xmin=1800, xmax=10200,
    ymin=0, ymax=4200,
    xtick={2000,4000,6000,8000,10000},
    ytick={0,1000,2000,3000,4000},
    scaled y ticks=base 10:-3,
    legend pos=north west,
    grid style=dashed,
    legend style={nodes={scale=0.6, transform shape}}
]
\addplot[
    color=olive, dotted, line width=0.4mm
    ] file {Data_Files/Mod_Ent_Wd_Int_brane_mu_1p3_y1_2000_y2_10000_d_4.dat};
\addplot[
    color=teal, line width=0.4mm
    ] file {Data_Files/Full_New_SubAd_d_4_mu_1p3_y1_2000_y2_10000.dat};
    \addplot[
    color=black!60!red,dashed,line width=0.4mm
    ] file {Data_Files/Const_Ext_Curv_alpha_1_mu_1p3_y1_2000_y2_10000_d_4.dat};
    \addplot[
    color=black!30!red,dashed,line width=0.4mm
    ] file {Data_Files/Const_Ext_Curv_alpha_p5_mu_1p3_y1_2000_y2_10000_d_4.dat};
    \addplot[
    color=red,dashed,line width=0.4mm
    ] file {Data_Files/Const_Ext_Curv_alpha_p1_mu_1p3_y1_2000_y2_10000_d_4.dat};
    \addlegendentry{$\partial^+\mathcal{H}(B)$};
    \addlegendentry{$\partial^+\mathcal{\Tilde{C}}(B)$};
    \addlegendentry{$\partial^+\mathcal{E}_{1}(B)$};
    \addlegendentry{$\partial^+\mathcal{E}_{0.5}(B)$};
    \addlegendentry{$\partial^+\mathcal{E}_{0.1}(B)$};
    
\end{axis}
\end{tikzpicture}
    \caption{Comparing $\mathcal{H}(B)$ and $\mathcal{E}_{\alpha}(B)$ and bound on $\mathcal{C}(B)$ ($y_1=2000$, $y_2=10000$) for $\mu_*=1.0$ (top panel) and $\mu_*=1.3$ (bottom panel). The AdS length scale $L$ and the bulk/boundary speed of light $c$ have been set to $1$.}
    \label{fig:DoD_Compare_3-4D}
\end{figure}
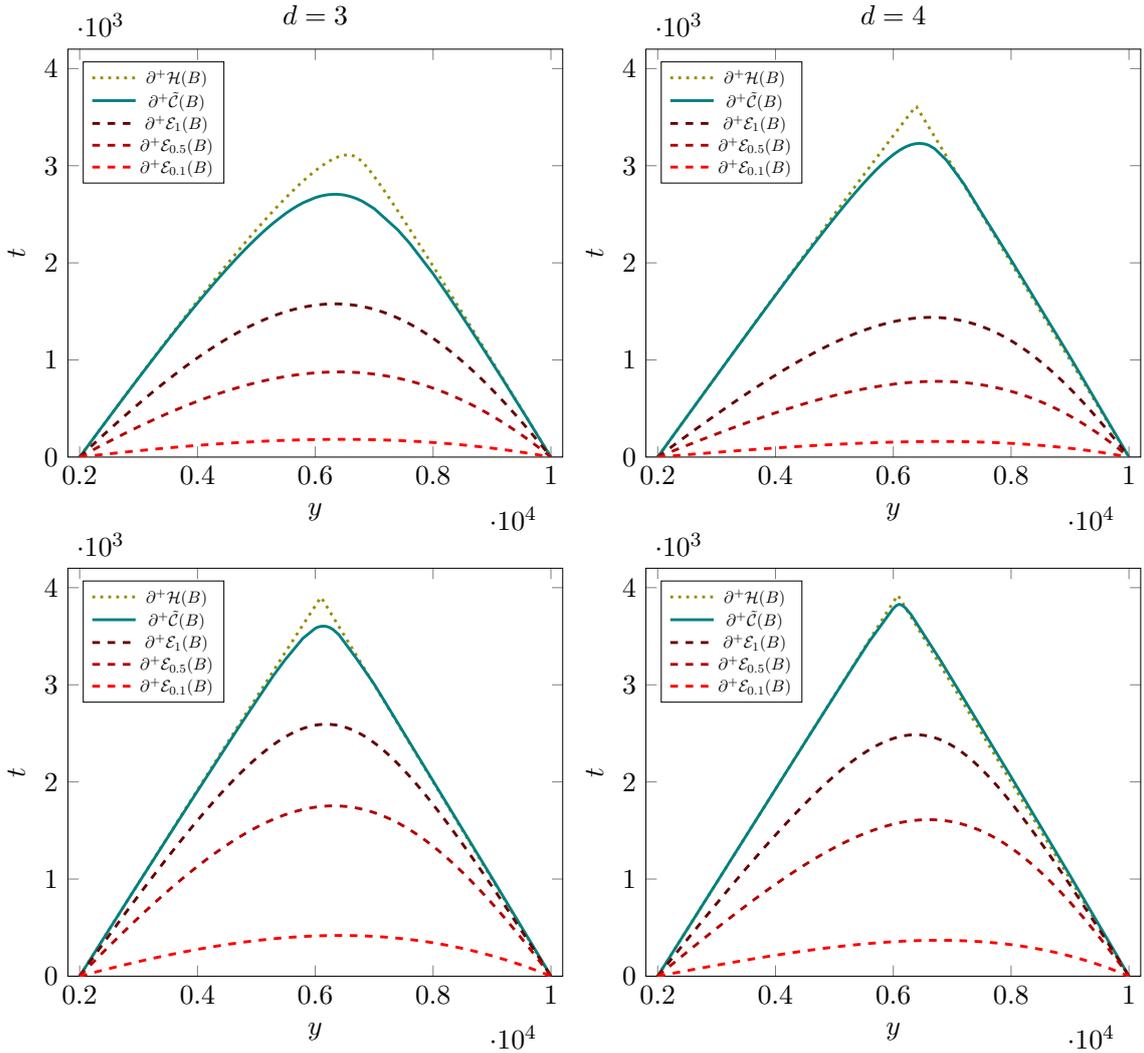
Let us now compare the different candidate domains of dependence with each other. The comparisons for strip regions on a $d=2$ dimensional brane is illustrated in \cref{fig:d2_DoD_Compapre} while the corresponding results for strips regions on $d=3$ and $d=4$ dimensional branes are shown in \cref{fig:DoD_Compare_3-4D}. The dotted lines illustrate the future boundary of the intersection of the entanglement wedge with the brane while the dashed lines mark the future boundary of the EFT domains of dependence. The upper bound on the unitary domains of dependence are shown as solid curves. 
As can be seen from the figures, the lines indicating the boundary of the EFT domain of dependence (i.e., the dashed curves) are contained in the bounds on $\mathcal C(B)$ and $\mathcal H(B)$ with $d=2$ being the limiting case.

These results imply that if the subregions we consider are part of a nice slice, then we remain in both the unitary and the holographic domains of dependence. Hence, if the nice slice criterion is necessary for an EFT description, there is no way to decide which of the unitary or the holographic domain of dependence is more fundamental. Further, as long as the subregions being considered are part of a nice Cauchy slice, subadditivity is not violated. Hence, violation of subadditivity, just like superluminal signalling (\cref{sec:scale_of_violation}) is purely a UV effect which is beyond the regime of predictability of EFT. In addition, our results indicate that the nice slice criterion could be too restrictive in dimensions $d \geq 2$, at least for the simple case of AdS vacuum studied here.

Note that the $d=2$ dimensional brane is special in many ways. The different domains of dependence are symmetric with respect to coordinate center of the partial Cauchy slice on the $d=2$ dimensional brane unlike in higher dimensions. For the $d=2$ dimensional brane, the bound on $\mathcal{C}(B)$ and the entanglement wedge-brane intersection coincide exactly with the boundary of the $\alpha=1$ EFT domain of dependence $\mathcal{E}_{\alpha=1}(B)$ (\cref{fig:d2_DoD_Compapre}). In other words, in the $d=2$ case, the SA bound and the boundary of the entanglement-brane intersection are curves of constant extrinsic curvature.

\section{Discussion}

In this paper we argued that the non-local effects discussed in \cite{Omiya:2021olc} are of UV origin and do not affect treating the brane perspective of double holography as an EFT which is local at energy scales below an energy cutoff scale $1/\Lcut$, roughly set by the bulk AdS radius. Moreover, we argued that the naive causal structure given by the metric induced on the brane is corrected. However, these corrections lie outside the usually assumed regime of validity of the effective brane description via the nice-slice criterion. Our results should be seen as evidence in favour of bottom-up braneworld models being self-consistent and good approximations to more complicated top-down constructions. As we have seen, merely realizing that the brane theory is an EFT with some cutoff scale is sufficient to avoid (at least some) obvious inconsistencies. 

However, this also means that when working with these models we need to be careful only to ask questions which can be answered within the EFT. Moreover, the range of validity of Karch-Randall models is by no means large. Depending on the location of the brane in the bulk, there are different effective speeds of light and different unitary and holographic domains of dependence, which can roughly be characterized by the location of the brane $\mu_*$.  One extreme example is the tensionless limit studied in \cite{Geng:2020qvw} corresponding to $\mu_* = 0$. Here, the domains we discussed reduce to a constant time slice on the brane, since the effective speed of light $c_\text{eff}$ diverges. In this limit the brane perspective simply does not make sense.

In \cref{sec:eft_and_subregion_entropies} we proposed that entanglement inequalities for generalized entropy might shine light on the applicability of effective field theory. We believe that it would be interesting to develop this idea further. For instance, one could imagine using other, stricter inequalities like strong subadditivity (SSA) to further constrain the class of partial Cauchy slices that allow for an EFT description.\footnote{Since there are more subregions involved in SSA, it is not completely obvious what relative configurations of Cauchy slices must be used to look for violations. In fact, we did observe SSA violation using symmetric configurations of strip regions on the $d=2$ dimensional brane and obtained exactly the same bound on the unitary domain of dependence as obtained from SA violation.}  
Of course, this is not the only way to study reliability of EFT in KR braneworlds. For example, another interesting direction pursued by another group is to investigate causality in braneworld models through performing quantum tasks in the brane perspective via scattering in the bulk \cite{Mori:2023swn}.

In the context of double holography, we defined unitary, holographic and EFT domains of dependence as candidate replacements for the notion of domain of dependence on the brane. The observation that the different candidate domains of dependence coincide on $d=2$ dimensional branes despite being defined from very different and independent motivations, is very intriguing. We do not completely understand the origin of this coincidence and it requires further analysis. Understanding this better could perhaps inform us about some potential relationships between their independent motivations, namely unitarity of the boundary theory, bulk reconstruction and the nice slice criterion in this setting. Further, it would be interesting to understand the relationship among the three regions better in arbitrary dimensions. This could perhaps shed light on how one should classify the \emph{true} domain of dependence, i.e., the instances in which we can predict the quantum state by time evolution, going beyond the naive geometric picture advocated for in this paper.

Further, a more in-depth study of the existing models of black hole evaporation and the regime of validity and predictability of the brane computations in those settings is interesting and also necessary. This is particularly relevant for approaches using branes with quantum black holes or the c-metric \cite{Emparan:2020znc, Emparan:2023dxm}. Moreover, it also seems worthwhile extending the computations to the case of de Sitter branes \cite{Iwashita:2006zj} or models with additional DGP couplings \cite{Dvali:2000hr,Chen:2020hmv} on the brane.

More generally, the precise conditions under which semi-classical gravity coupled to matter is valid are still unknown and we are hopeful that some of the insights and techniques from double holography can inform a more general approach to map out the limits of semi-classical gravity.
In this context it seems interesting to investigate if one can find regions within double holography where the EFT domain of dependence is not completely contained  in the unitary or holographic domains of dependence 
for moderate values of $\alpha \sim \mathcal O(0.1)$ or smaller. This might indicate the breakdown of semi-classical gravity in regions where we naively expect it to be true.

A crucial ingredient in the discussion of \cref{sec:eft_in_double_holo} was that generalized entropy in double holography can be computed by HRT surfaces anchored to the brane. While this prescription is accepted in the case where HRT surfaces connect the asymptotic boundary and the brane, we have argued here and elsewhere \cite{Neuenfeld:2021bsb} why this statement should hold more generally. However, a proof of the HRT formula on the brane is missing. One might therefore worry that, e.g., for subregions which are highly boosted, the HRT surfaces fail to produce the correct generalized entropy and thus the violation of subadditivity found above are nothing but an artifact. However, even when taking this pessimistic view, our results of \cref{sec:eft_in_double_holo} show that the HRT formula defines a consistent measure of entropy in regions where we believe the EFT to be valid based on other reasoning, such as the nice slice argument.

Relatively little of our arguments and computations rely on double holography and we expect most of the discussion to carry over to other models of cutoff holography, e.g., to $T\bar T$ deformations of holographic CFTs without any major modifications \cite{McGough:2016lol}. $T\bar T$ deformations share some properties with the brane description in double holography. Most notably it is conjectured that deforming a holographic CFT$_2$ with its stress energy tensor moves the asymptotic boundary into the bulk. Consequently, many effects such as violation of subadditivity, superluminal signalling 
and emergence of non-local physics can also be observed in this setup. In particular, note that entanglement entropy in $T \bar T$ is conjectured to be computed by HRT surfaces anchored on the cutoff surface \cite{Donnelly:2018bef}. It would therefore be interesting to revisit discussions of entanglement wedges in $T\bar T$ with this in mind \cite{Lewkowycz:2019xse}. Interestingly, the subadditivity bound on the unitary domain of dependence and the holographic domain of dependence, two independently defined candidate replacements for domain of dependence coincide exactly for the case relevant for $T\Bar{T}$ deformations ($d=2$ dimensional branes),  hinting at some unexplored connection between unitarity of the boundary theory and bulk reconstruction in such setups. There are, however, also important differences between $T\Bar{T}$ deformations and double holography. The boundary condition at the cutoff surface in $T \bar T$ is usually taken to be Dirichlet (see however \cite{Guica:2019nzm}). Moreover, the UV completion in double holography is given by a BCFT and it is unclear if there is any relation to $T \bar T$ deformations. It would therefore be interesting to better understand the relation between $T \bar T$ deformations and Karch-Randall-Sundrum models.

Lastly, apart from the Karch-Randall model, similar superluminal signalling has also been reported in a number of cosmological braneworld models, especially in the contexts of finding a non-inflationary solution to the cosmological horizon problem \cite{Chung:1999xg,Abdalla:2002ir,Abdalla:2004xm,Greene:2022urm, Dai:2023zsx}, finding signs for the existence of extra compactified dimensions \cite{Polychronakos:2022uam} and
exploring the capacity for real time communication across arbitrarily large distances \cite{Greene2}. It would be interesting to investigate the implications of the EFT reasoning advocated for in this paper in such cosmological settings.

\acknowledgments
\label{sec:acknowledgements}
The authors would like to extend special thanks to Rob Myers for his inputs during the initial stages of this work and for several insightful discussions. The authors also thank Chris Akers, Luis Apolo, Jan de Boer, Soumangsu Chakraborty, Ben Freivogel, Andreas Karch, Adam Levine, Alex May, Takato Mori, Andrew Rolph, Kamran Salehi Varizi, Krishan Saraswat, Erik Verlinde, Masataka Wanatabe, Beni Yoshida for interesting conversations related to this work. During this work DN was partially supported by the Simons Foundation through the ``It from Qubit'' collaboration and the Heising-Simons Foundation “Observational Signatures of Quantum Gravity” collaboration grant 2021-2817. MS thanks the Center for Theoretical Physics, Department of Physics, MIT for support through the First Year Graduate Fellowship award and the Perimeter Institute for support through the Perimeter Scholars International Scholarship award. Initial stages of this work was done at Perimeter Institute for Theoretical Physics, Canada. Research at Perimeter Institute is supported in part by the Government of Canada through the Department of Innovation, Science and Economic Development Canada and by the Province of Ontario through the Ministry of Colleges and Universities.

\appendix

\section{Brane Subregion Entropies in 2+1 Dimensions}
\label{sec:2+1_entropy}
In $2+1$ dimensions, a strip region $B$ is simply a one-dimensional subregion on the boundary cylinder. To get the HRT surface associated with it, we need to extremize the functional
\begin{equation}
\label{eqn:time-dep-extremal-length}
    \Delta s=\int \frac{\sqrt{z'^2+x'^2-t'^2}}{z}d\lambda\equiv\int L_t\,d\lambda,
\end{equation}
where $t'\equiv\frac{dt}{d\lambda}$, $z'\equiv\frac{dz}{d\lambda}$ and $x'\equiv\frac{dx}{d\lambda}$. The boundary conditions are that the extremal surface must pass through the end points of $B$. The Euler-Lagrange-equations are
\begin{equation}
\label{eqn:t-dep-x-eq}
    \frac{\partial L_t}{\partial x}=\frac{d}{d\lambda}\left(\frac{\partial L_t}{\partial x'}\right)\implies\frac{x'}{z\sqrt{x'^2+z'^2-t'^2}}=C_x,
\end{equation}
\begin{equation}
\label{eqn:t-dep-t-eq}
    \frac{\partial L_t}{\partial t}=\frac{d}{d\lambda}\left(\frac{\partial L_t}{\partial t'}\right)\implies\frac{t'}{z\sqrt{x'^2+z'^2-t'^2}}=C_t,
\end{equation}
with constants $C_x, C_t$.
Similarly, the $z$ Euler-Lagrange-equation gives:
\begin{equation}
\label{eqn:t-dep-z-eq}
  \frac{-t'^4+t'^2 \left(2 x'^2+z z''+z'^2\right)-z t' t'' z'-x' \left(-z x'' z'+x'^3+x' \left(z z''+z'^2\right)\right)}{z^2 \left(-t'^2+x'^2+z'^2\right)^{3/2}}=0.
\end{equation}
We observe that one of the Euler Lagrange equations is redundant. Substituting for $x'$, $x''$, $t'$, $t''$ from \cref{eqn:t-dep-x-eq}, \cref{eqn:t-dep-t-eq} in \cref{eqn:t-dep-z-eq} makes it trivially zero. Hence any simultaneous solution of \cref{eqn:t-dep-x-eq,eqn:t-dep-t-eq} will automatically satisfy \cref{eqn:t-dep-z-eq}. This implies that we need to only solve the $x$ and $t$ equations to solve the complete system of equations. The $x$ and $t$ equations can be formulated as
\begin{equation}
\label{eqn:t-x_after_massage}
    \left(\frac{dt}{dx}\right)=\frac{C_t}{C_x} \qquad \text{and} \qquad
    \left(\frac{dz}{dx}\right)^2=\frac{1}{C_x^2z^2}-\left(1-\left(\frac{C_t}{C_x}\right)^2\right).
\end{equation}
The solutions to these equations depend of the relative magnitude of $C_t$ and $C_x$.
\begin{enumerate}
    \item For the $C_x^2=C_t^2$ case the solutions are
\begin{equation}
    \frac{dz}{dx}=\pm\frac{1}{C_x z}\implies x=\pm\frac{C_x}{2}z^2+D_1
\end{equation}
\begin{equation}
    \frac{dt}{dx}=\frac{C_t}{C_x}\implies t=\frac{C_t}{C_x}x+D_2,
\end{equation}
where $D_1$ and $D_2$ are integration constants. Using this solution to compute the extremal length/area from \cref{eqn:time-dep-extremal-length}, we get
\begin{equation}
\label{eqn:diff_timeslice_CxCtsame}
    \Delta s=\int\frac{z'}{z}d\lambda=\int_{z_1}^{z_2}\frac{dz}{z}=\log(z_2)-\log(z_1),
\end{equation}
where $z_2$ and $z_1$ are the $z$-coordinates of the end points of $B$. To get the length, we need evaluate the above expression such that $z_2>z_1$. 

    \item Now moving on to the case of $C_x^2\neq C_t^2$. For this case the solutions to \cref{eqn:t-dep-t-eq,eqn:t-dep-x-eq} are given by 
\begin{equation}
\label{eqn:t-dep_parametric_sol_1}
    \frac{dt}{dx}=\frac{C_t}{C_x}\implies t=\frac{C_t}{C_x}x+D_3
\quad \text{and} \quad
    \frac{(x+d_1)^2}{\frac{C_x^2}{(C_t^2-C_x^2)^2}}+\frac{z^2}{\frac{1}{(C_x^2-C_t^2)}}=1,
\end{equation}
where $d_1$ and $D_3$ are integration constants. The solution can be parametrized in terms of $\lambda$ as
\begin{equation}
\label{eqn:t-dep_parametric_sol_new_1}
    x=\frac{C_x}{C_x^2-C_t^2}\cos{\lambda}-d_1 \quad,\quad z=\frac{1}{\sqrt{C_x^2-C_t^2}}\sin{\lambda}~\rightarrow \text{for}~C_x^2>C_t^2
\end{equation}
\begin{equation}
\label{eqn:t-dep_parametric_sol_new_2}
    x=\frac{C_x}{C_t^2-C_x^2}\cosh{\lambda}-d_1\quad,\quad z=\frac{1}{\sqrt{C_t^2-C_x^2}}\sinh{\lambda}~\rightarrow \text{for}~C_t^2>C_x^2.
\end{equation}
Using the solutions to compute the extremal length/area from \cref{eqn:time-dep-extremal-length}, we get:
\begin{equation}
    \Delta s=\int^{\lambda_2}_{\lambda_1}\frac{d\lambda}{\sin{\lambda}}=\log\left(\frac{\tan{\frac{\lambda_2}{2}}}{\tan\frac{\lambda_1}{2}}\right),~\qquad \text{for}~C_x^2>C_t^2,
\end{equation}
\begin{equation}
\label{eqn:diff_timeslice_CtGCx}
    \Delta s=\int^{\lambda_2}_{\lambda_1}\frac{d\lambda}{\sinh{\lambda}}=\log\left(\frac{\tanh{\frac{\lambda_2}{2}}}{\tanh\frac{\lambda_1}{2}}\right),~\qquad \text{for}~C_t^2>C_x^2,
\end{equation}

where $\lambda_1$ and $\lambda_2$ (with $\lambda_2>\lambda_1>0$) are the parameter values at the endpoints of $B$ in our description. Thus, in order to evaluate the entropy, we only need to get hold of $\lambda_2$ and $\lambda_1$. In the way we have set up the calculations there are six unknowns - the integration constants and the $\lambda$'s at the end points ($d_1$, $D_3$, $C_x$, $C_t$, $\lambda_1$ and $\lambda_2$). What we know are the coordinates ($t$, $z$, $x$) of the endpoints of $B$ (by virtue of the choice of $B$). To get the extremal surface for the sub-region $B$, we have to impose the boundary conditions that the parameterized solutions (\cref{eqn:t-dep_parametric_sol_new_1}, \cref{eqn:t-dep_parametric_sol_new_2}) pass through the two endpoints of $B$. Equating the $t$, $x$ and $z$ coordinates at the 2 endpoints give two equations each. These are a total of six equations that can be used to fix the six unknowns. Note that the most relevant case for defining states on partial Cauchy slices on a $d=2$ dimensional brane is when $C_x^2>C_t^2$, i.e., when $B$ is `spacelike'. But the other cases do become important in higher dimensions. as discussed in the main text.
\end{enumerate}
\section{Intersection of Entanglement Wedges and the Brane in 2+1 Dimensions}
\label{sec:Ent_intersection_brane}
\begin{figure}
    \centering
    \includegraphics[scale=0.4]{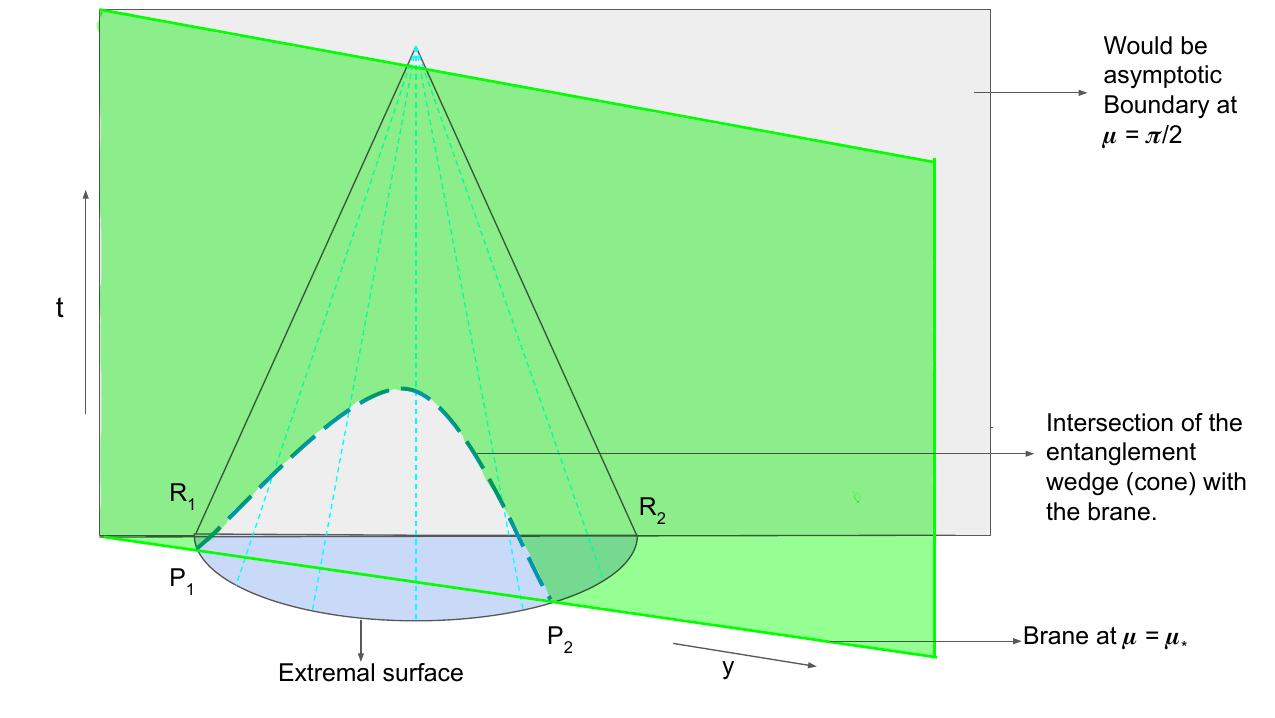}
    \caption{Illustrating intersection of the entanglement wedge with the brane}
    \label{fig:Brane_int_entWed}
\end{figure}
Since we have the analytical expressions for the HRT surfaces, \cref{eqn:t-dep_parametric_sol_1}, in the $d=2$ case, computing the entanglement wedge $\mathcal{W_E}[B]$ associated to a constant (Poinca\'e) time partial Cauchy slice $B$ is quite straightforward. The extremal surface associated to a subregion $B$ on the brane can be thought to be a part of an extremal surface associated with some other subregion $\Tilde{B}$ anchored on the imagined asymptotic boundary beyond $\mu=\mu_*$ at $R_1(t=0,z=0,x=x_1)$ and $R_2(t=0,z=0,x=x_2)$ (recall the Poincar\'e coordinates \cref{eq:poincare} and see \cref{fig:Brane_int_entWed}). Since the extremal surface associated to $\Tilde{B}$ is semicircular (\cref{eqn:t-dep_parametric_sol_1} with $C_t=0$), $\mathcal{W_E}[\Tilde{B}]$ (obtained by shooting null rays from the semicircle to the imagined asymptotic boundary)
 would just be a double cone. 
$\mathcal{W_E}[B]$ then is the region enclosed between the brane and the double cone surface with the two intersecting in the $t=0$ plane at $P_1$ and $P_2$, which are the endpoints of $B$. Since we have been looking only at regions with $t>0$ in our attempt to bound $\mathcal{C}(B)$, we will only consider the intersection of the upper half of the double cone with the brane. See \cref{fig:Brane_int_entWed} for a visualization. The equation of the surface of the upper half of the double cone is given by
 \begin{equation}
     z^2+\left(x-\frac{x_1+x_2}{2}\right)^2=\left(\frac{x_2-x_1}{2}-t\right)^2.
 \end{equation}
We can substitute for $z$ and $x$ in terms of $y$ and $\mu$ using \cref{eqn:Slicing_to_poincare}. The intersection of the surface of the cone with the brane is given by setting $\mu=\mu_*$,
\begin{equation}
\label{eqn:brane_intersect_ent_wedge}
    y^2\cos^2\mu_*+\left(y\sin\mu_*-\frac{x_1+x_2}{2}\right)^2=\left(\frac{x_2-x_1}{2}-t\right)^2.
\end{equation}
 \begin{figure}
    \centering
    \includegraphics[scale=0.75]{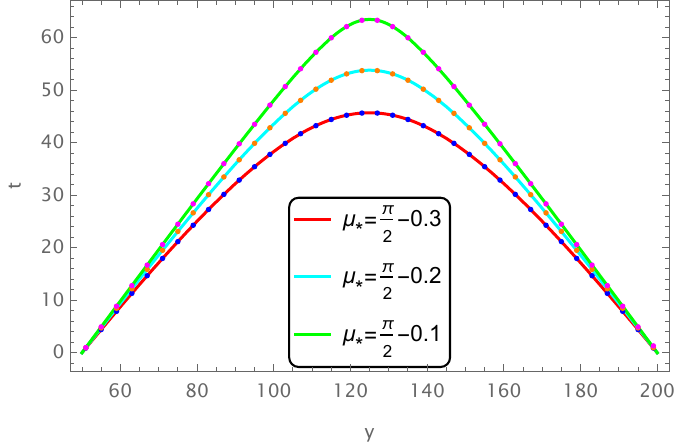}
    \caption{The dotted lines are the boundary of the region in which sub-additivity holds, i.e., the bound on $\mathcal{C}(B)$.  $y_1=50$, $y_2=200$ and the solid lines represent the curves traced by \cref{eqn:final_intersection_eq} for different $\mu_*$'s. }
    \label{fig:Sub-Str-Ent_Agreement}
\end{figure}
This expression contains $x_1$ and $x_2$ which we need to write in terms of $y_1$ and $y_2$ in order to compare with the region of validity of entropic inequalities that we computed for some fixed $y_1$ and $y_2$. Since $P_1$ and $P_2$ are points at which the brane intersects the cone surface in the $t=0$ plane, $y_1$ and $y_2$ are the two roots of \cref{eqn:brane_intersect_ent_wedge} when we set $t=0$. Choosing $P_1$ as the point closer to $R_1$ and $P_2$ as the point closer to $R_2$, we get
 \begin{equation}
 \label{eqn:x1x2-y1y2}
 \begin{split}
     & x_1(y_1,y_2)= \frac{1}{2} \csc (\mu_* ) \left(y_1+y_2-\sqrt{y_1^2+2 y_1 y_2 \cos (2 \mu_* )+y_2^2}\right),\\
     & x_2(y_1,y_2)= \frac{1}{2} \csc (\mu_* ) \left(y_1+y_2+\sqrt{y_1^2+2 y_1 y_2 \cos (2 \mu_* )+y_2^2}\right).
 \end{split}
 \end{equation}
 Now we have the equation for the intersection of $\mathcal{W_E}[B]$ with the brane
 for $t>0$,
 \begin{equation}
 \label{eqn:final_intersection_eq}
 \begin{split}
     t &=\frac{x_2(y_1,y_2)-x_1(y_1,y_2)}{2}-\sqrt{y^2\cos^2\mu_*+\left(y\sin\mu_*-\frac{x_1(y_1,y_2)+x_2(y_1,y_2)}{2}\right)^2}\\
     & =\frac{1}{2\sin\mu_*}\left(\sqrt{y_1^2+y_2^2+2y_1y_2\cos(2\mu_*)}-\sqrt{y^2\sin^2(2\mu_*)+\left(y_1+y_2-2y\sin^2\mu_*\right)^2}\right).
 \end{split}
 \end{equation}
 This is actually a constant extrinsic curvature strip region on the brane with $\mathfrak{K}=\sin\mu_*/L$ and coincides exactly with the subadditivity bound on the unitary domain of dependence $\mathcal{C}(B)$ as seen in \cref{fig:Sub-Str-Ent_Agreement}.


\bibliographystyle{JHEP}
\bibliography{references.bib}

\end{document}